\documentclass[aip,graphicx]{revtex4-1}
\usepackage[cp1251]{inputenc}
\usepackage[T2A]{fontenc}
\usepackage[english]{babel}
\usepackage{amssymb,latexsym,amsmath,amscd}
\usepackage{graphicx,color,framed}
\usepackage{xcolor}
\usepackage{siunitx} 
\begin{document}
%\selectlanguage{russian}
\selectlanguage{english}
\title{Statistical field theory of ion-molecular solutions}
\author{\firstname{Yu.A.} \surname{Budkov}}
\email[]{ybudkov@hse.ru}
%\homepage[]{Your web page}
%\thanks{}
%\altaffiliation{}
\affiliation{School of Applied Mathematics, National Research University Higher School of Economics, Tallinskaya st. 34, 123458 Moscow, Russia}
\affiliation{G.A. Krestov Institute of Solution Chemistry of the Russian Academy of Sciences, 153045, Akademicheskaya st. 1, Ivanovo, Russia}

\begin{abstract}
In this article, I summarize my theoretical developments in the statistical field theory of salt solutions of zwitterionic and multipolar molecules. Based on the Hubbard-Stratonovich integral transformation, I represent configuration integrals of dilute salt solutions of zwitterionic and multipolar molecules in the form of functional integrals over the space-dependent fluctuating electrostatic potential. In the mean-field approximation, for both cases, I derive integro-differential self-consistent field equations for the electrostatic potential, generated by the external charges in solutions media, which generalize the classical Poisson-Boltzmann equation. Using the obtained equations, in the linear approximation, I derive for the both cases a general expression for the electrostatic potential of a point-like test ion, expressed through certain screening functions. I derive an analytical expression for the electrostatic potential of the point-like test ion in a salt zwitterionic solution, generalizing the well known Debye-Hueckel potential. In the salt-free solution case, I obtain analytical expressions for the local dielectric permittivity around the point-like test ion and its effective solvation radius. For the case of salt solutions of multipolar molecules, I find a new oscillating behavior of the electrostatic field potential of the point-like test ion at long distances, which is caused by the nonzero quadrupole moments of the multipolar molecules. I obtain a general expression for the average quadrupolar length of a multipolar solute. Using the random phase approximation (RPA), I derive general expressions for the excess free energy of bulk salt solutions of zwitterionic and multipolar molecules and analyze the limiting regimes resulting from them. I generalize the salt zwitterionic solution theory for the case when several kinds of zwitterions are dissolved in the solution. In this case, within the RPA, I obtain a general expression for the solvation energy of the test zwitterion. Finally, I demonstrate how to take a systematic account of the excluded volume correlations between multipolar molecules in addition to their electrostatic correlations. I believe that the formulated findings could be useful for the future theoretical models of the real ion-molecular solutions, such as salt solutions of micellar aggregates, metal-organic complexes, proteins, betaines, {\sl etc.}
\end{abstract}
\maketitle

\section{Introduction}
Coarse-grained modeling (CGM) methods have been so far widely used for molecular dynamics (MD) simulations of equilibrium properties of ion-molecular systems, such as ionic and zwitterionic liquids, dielectric polymers, micellar aggregates, proteins, and complex colloids (see, for example, \cite{wang2007understanding,bhargava2007nanoscale,shinoda2010zwitterionic,chakrabarty2010coarse,baaden2013coarse,nielsen2004coarse,wu2011coarse,schroder2010simulating,cavalcante2014polarizability}). CGM is based on the idea that separate atomic (molecular) groups are combined into large particles, whose interactions are described by effective pairwise potentials. Despite a certain success in reducing the computing time of the MD simulation by coarsening of the real structure of the molecules in comparison with their full-atomic representation, modeling of the systems contaning a large number of particles (even with a coarsened structure) under the conditions of a condensed phase is still rather computationally expensive. The structural  properties of ion-molecular systems (for instance, site-site pair correlation functions and hydration numbers) are commonly described by the statistical theory of integral equations for the site-site correlation functions (Reference Interactions Site Model or RISM) with different closing relations (see, for instance, \cite{fedotova2015ion,fedotova2016proline,eiberweiser2015hydration,fedotova2019features,fedotova2014ion,fedotova20121d}). Although the RISM theory allows quite accurate estimation of the structural quantities, it is impossible within the RISM theory to obtain analytical expressions for the thermodynamic properties, such as the excess chemical potentials of species, osmotic coefficient, equation of state, {\sl etc.} \cite{ratkova2015solvation,kovalenko2004three,chuev2006quasilinear,kovalenko2005molecular}. Moreover, the numerical solution of the 3D-RISM integral equations is rather computationally expensive and no less time-consuming than the MD simulations.

Thus, it can be concluded that modern theoretical chemical physics and physical chemistry need some computational tools based on the fundamentals of statistical physics and allowing one to predict the qualitative behavior of the thermodynamic parameters in a fast and effective way. The field-theoretic (FT) methods based on the Hubbard-Stratonovich (HS) integral transformation of the configuration integral and correlation functions of a certain model fluid in the form of a functional integral over one or few space-dependent fluctuating fields could become such a tool. At present, these methods are well-proven as the ones for describing the thermodynamic and structural properties of simple neutral and ionic fluids \cite{Storer,Frusawa2018,IvanLis,hubbard,Caillol1,Caillol2,Caillol3,Edwards,efimov1996partition,zakharov1999classical,Brill:98,Trokhymchuk,brilliantov2020molecular} and polymer solutions \cite{edwards1966theory,fredrickson2006equilibrium}. Among them are the mean-field (saddle-point) approximation \cite{Brill:98,abrashkin2007dipolar,budkov2015modified,brilliantov2020molecular}, random-phase approximation \cite{edwards1966theory,budkov2018nonlocal,budkov2019Astatistical}, many-loop expansion \cite{netz2001electrostatistics}, variation method \cite{lue2006variational}, and the renormalization group theory \cite{brilliantov1998peculiarity}. However,  to  apply  the  FT  approaches  to  the theoretical description of ion-molecular systems at the equilibrium in the bulk and at the interfaces, one needs to take into account not only the universal dispersion and excluded volume inter-molecular interactions, but also the special features of the species molecular structure, such as multipole moments, electronic and molecular polarizabilities, configuration asymmetry of the molecule, {\sl etc}. However, scientists have only made the first steps in that direction. Up to now, they have formulated FT models of polarizable polymer solutions \cite{martin2016statistical,kumar2009theory,budkov2017polymer,budkov2017flory,gordievskaya2018interplay,kumar2014enhanced,grzetic2018effective,gurovich1994microphase}, salt solutions of polar and multipolar molecules \cite{budkov2018nonlocal,budkov2019Astatistical,budkov2019statistical,budkov2019nonlocal}, electrolyte solutions with an explicit account of the polar solvent and ion polarizability \cite{buyukdagli2013microscopic,buyukdagli2014dipolar,buyukdagli2013alteration,abrashkin2007dipolar,coalson1996statistical,budkov2015modified,grzetic2019contrasting}, solutions of liquid crystalline ionic fluids \cite{lue2006variational,kopanichuk2016steric}, aqueous solutions of the intrinsically disordered proteins \cite{lin2017random,mccarty2019complete}, and a cluster model of ionic liquids \cite{avni2020charge}. Despite the evident success achieved in applications of the FT methods to complex ion-molecular systems, they still remain underestimated by chemical engineers and materials scientists, in comparison with the coarse-grained MD simulation methods or the above mentioned RISM theory. In my opinion, this is because of their mathematical complexity, on the one hand, and the lack of accessible software for general public users, on the other hand.

Recent advances in experimental studies of the organic zwitterionic compounds (see, for instance, \cite{canchi2013cosolvent,heldebrant2010reversible,felder2007server}) consisting simultaneously of positively and negatively charged ionic groups, located at rather long distances (several nanometers) offer a challenge to statistical thermodynamics to develop analytical approaches for describing thermodynamic properties of such systems. As a rule, the fundamental difference of zwitterions from classical small polar molecules, such as water and alcohols, is that it is impossible to model them as particles carrying point-like dipoles \cite{onsager1936electric,kirkwood1939dielectric,ho1976statistical}. Instead, they must be considered as pairs of bonded oppositely charged ionic groups located at fixed or fluctuating distances from each other, thus supplementing the theory with an additional length scale. The latter is attributed to the effective distance between the charged groups. Moreover, since in practice such molecules are dissolved in a low-molecular weight polar solvent (primarily in water) in the presence of salt ions, it is necessary to develop a model of electrolyte solutions with a small additive of polar particles described as ionic pairs with fluctuating distances between the charged groups. Such a $"$nonlocal$"$ theory could solve a number of fundamental problems of the polar fluids theory. Firstly, such theory will allow one to describe the behavior of the point-like charge potential in an electrolyte solution medium with an addition of polar particles at the scale of their effective size. Secondly, this theory must be devoid of the nonphysical unavoidable divergence of the electrostatic free energy of the polar fluid with point-like polar molecules, which is the case in the existing $"$local$"$ theories (see, for instance, \cite{dean2012ordering,levy2012dielectric,budkov2016polarizable,budkov2017polymer}). A generalization of the dipolar particles model, which could be relevant to the modern physical chemistry of solutions, is the model of $"$hairy$"$ particles \cite{budkov2019Astatistical}. Each hairy particle is comprised of a big charge, located at the center of the particle (central charge), and peripheral charges, located at fixed or fluctuating distances around it. Such configurations could be realized for spherical polyelectrolyte stars and brushes (see, for instanse, \cite{ballauff2007spherical,jusufi2002counterion}), spherical colloidal particles with counterions condensed onto their charged surfaces \cite{linse1999electrostatic,linse2005simulation}, metal-organic complexes with multivalent metal ions, surrounded by oppositely charged monovalent ionic ligands \cite{xu2014recent,perez2008stable}, {\sl etc.}

To the best of my knowledge, until recently, there were no attempts to formulate statistical models of the salt solutions of dipolar and multipolar molecules making it possible to obtain the analytical relations for their thermodynamic quantities from the first principles of statistical mechanics. In this regard, in the present extended article, in the context of modern statistical field theory of ion-molecular solutions, I would like to summarize my recent theoretical findings in the development of the FT models of ion-molecular solutions, whose solute molecules possess a multipolar internal electric structure.

\section{Statistical field theory of salt zwitterionic solutions}
The existing statistical theories of polar fluids describe molecules as point-like dipoles \cite{abrashkin2007dipolar} or hard spheres with a point-like dipole in their center \cite{ho1976statistical}. As is well known, disregard of the details of the internal electrical structure of polar molecules in calculations of the electrostatic free energy of the polar fluid from the first principles of statistical mechanics leads to $"$ultraviolet divergence$"$ of electrostatic free energy, which makes the researchers resort to the artificial $"$ultraviolet cut-off$"$, while integrating over the vectors of the reciprocal space \cite{levy2012dielectric,budkov2017polymer}. On the other hand, as it has been mentioned above, a $"$local$"$ theory that does not consider the charge distribution of the polar molecule, cannot describe the behavior of the point-like charge potential surrounded by polar molecules at the distances of their linear size order. Thus, one can expect that a $"$non-local$"$ statistical theory taking such details into account will be free from the ultraviolet divergence of the free energy present in the local theories and will allow one to investigate the electrostatic potential of the point-like ion at the distances from the ion comparable with the characteristic distances between the charged centers of polar molecules.

Let us consider an electrolyte solution comprised of a mixture of point-like ions of $s$ kinds with the total numbers $N_{i}$ ($i=1,...,s$) and charges $z_{i}e$ ($z_{i}$ is the ion valency and $e$ is the elementary charge) and $N$ zwitterions with point-like ionic groups with charges $\pm q$, confined in the volume $V$ at the temperature $T$. Let the number of ions $N_{i}$ satisfy the electrical neutrality condition $\sum_{i} z_{i}N_{i}=0$.
Moreover, let us assume that all the particles are immersed in a solvent, which is modelled as a continuous dielectric medium with the constant permittivity $\varepsilon$. This means the adoption of the simplest implicit solvent model which does not take into account the molecular structure effects. It can be justified by the assumption that the effective size of a solvent molecule is much smaller than the zwitterion effective size. The models of polar solvents taking into account the molecular structure were formulated in papers \cite{kornyshev1996shape,leikin1990theory,kornyshev1989nonlocal,kornyshev1986non,buyukdagli2014dipolar,buyukdagli2013alteration,buyukdagli2013microscopic}. I suppose that there is a probability distribution function \footnote{In integral equations theory of molecular liquids the probability distribution function $\omega(r)$ is usually called as intramolecular correlation function.} $\omega(\bold{r})$ of the distance between the charged centres, associated with each zwitterion. I would also like to note that the contribution of the repulsive forces between all the particles attributed to their excluded volume is neglected. It can be justified by rather small concentrations of ions and polar particles. At the end of this section, I will show how it is possible to extend the formulated statistical field theory to account for the excluded volume interactions. 

\begin{figure}[h!]
\center{\includegraphics[width=0.6\linewidth]{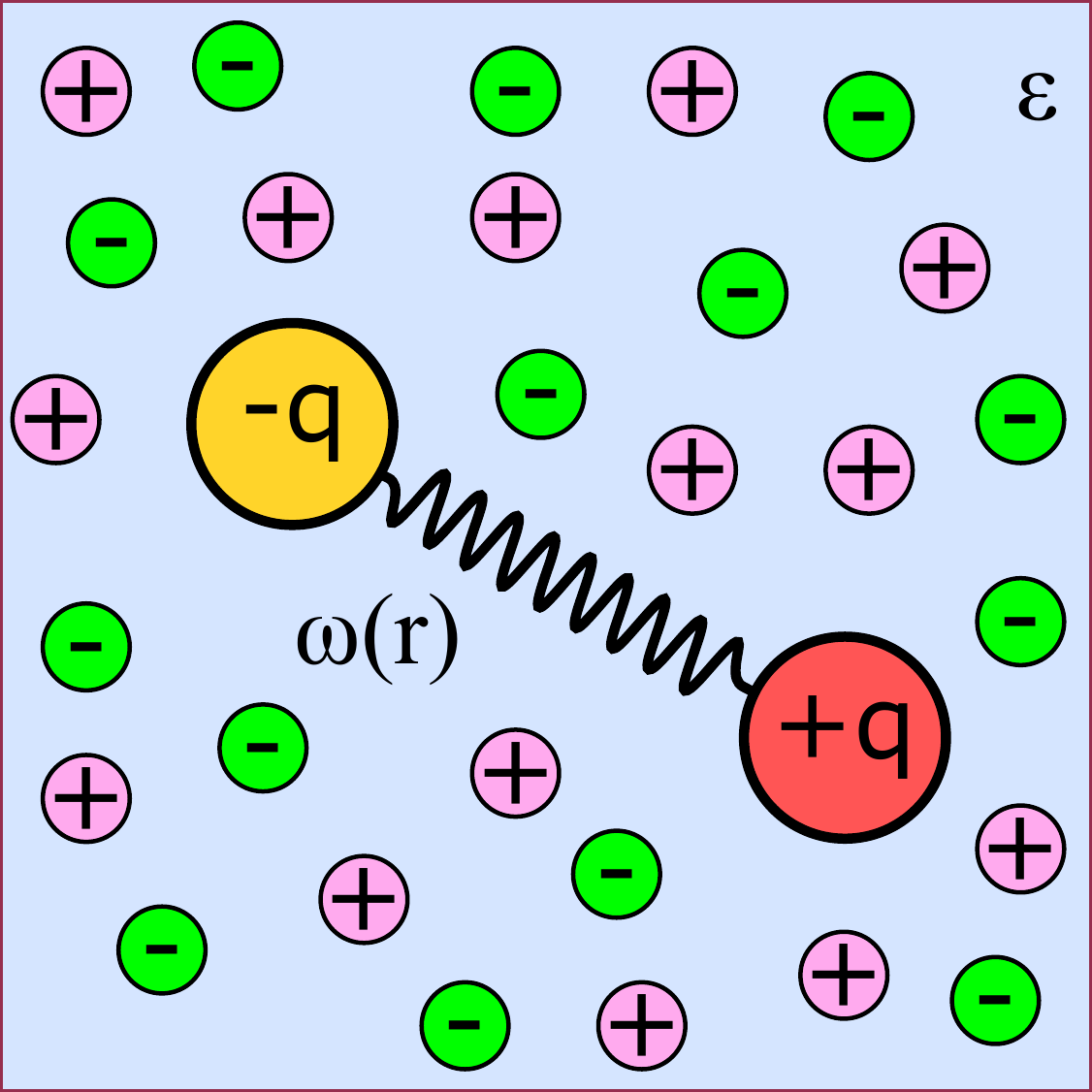}}
\caption{Schematic representation of the zwitterion surrounded by salt ions in a dielectric solvent medium.}
\label{potential}
\end{figure}

\textbf{General field-theoretic formalism.} In view of above-mentioned model assumptions, the configuration integral of the mixture can be written in the following form:
\begin{equation}
Q=\int d\Gamma_{zw}\int d\Gamma_s \exp\left[-\beta H_{int}\right],
\end{equation}
where
\begin{equation}
\int d\Gamma_{zw}(\cdot)=\int...\int\prod\limits_{j=1}^{N} \frac{d\bold{r}_j^{(+)}d\bold{r}_j^{(-)}}{V}\omega\left(\bold{r}_j^{(+)}-\bold{r}_j^{(-)}\right)(\cdot)
\end{equation}
is the measure of integration over the coordinates $\bold{r}_j^{(\pm)}$ of the ionic groups with the above-mentioned probability distribution function $\omega(\bold{r})$, which, according to its definition, must satisfy the normalization condition:
\begin{equation}
\int \omega(\bold{r}) d\bold{r}=1,
\end{equation}
and
\begin{equation}
\label{salt_int_meas}
\int d\Gamma_s (\cdot)=\int...\int\prod\limits_{j=1}^{s}\prod\limits_{k_{j}=1}^{N_{j}}\frac{d\bold{r}_{k_{j}}}{V}(\cdot)
\end{equation}
is the integration measure over the coordinates of the salt ions; $\beta = (k_{B}T)^{-1}$, $k_{B}$ being the Boltzmann constant. The total potential energy of the interactions taking into account the self-interactions of the charges is
\begin{equation}
\label{hamilt_el1}
H_{int}=\frac{1}{2}\int d\bold{r}\int d\bold{r}'\hat\rho(\bold{r})G_0(\bold{r}-\bold{r}')\hat\rho(\bold{r}')=\frac{1}{2}\left(\hat\rho G_0 \hat\rho\right),
\end{equation}
where $G_0(\bold{r}-\bold{r'})=1/(\varepsilon|\bold{r}-\bold{r}'|)$ is the Green function of the Poisson equation and
\begin{equation}
\hat\rho(\bold r)=\hat{\rho}_{zw}(\bold{r})+\hat{\rho}_{i}(\bold{r})+\rho_{ext}(\bold r),
\end{equation}
is the total charge density of the system with the microscopic charge densities of the zwitterions
\begin{equation}
\hat{\rho}_{zw}(\bold{r})=q\sum_{j=1}^{N}\left(\delta\left(\bold{r}-\bold{r}_j^{(+)}\right)-\delta\left(\bold r-\bold{r}_j^{(-)}\right)\right),
\end{equation}
and the ionic species
\begin{equation}
\label{ch_dens_ion}
\hat{\rho}_{i}(\bold{r})=e\sum\limits_{j=1}^{s}z_{j}\sum\limits_{k_{j}=1}^{N_{j}}\delta\left(\bold{r}-\bold{r}_{k_{j}}\right),
\end{equation}
and $\rho_{ext}(\bold{r})$ is the charge density of the external charges (of electrodes or membranes, for instance); $e$ is the elementary charge.

Using the standard HS transformation \cite{Stratonovich,Hubbard1,efimov1996partition}
\begin{equation}
\label{HS}
\exp\left[-\frac{\beta}{2}(\hat\rho G_0\hat\rho)\right]=\int\frac{\mathcal{D}\varphi}{C}\exp\left[-\frac{\beta}{2}(\varphi G_0^{-1}\varphi)+i\beta(\hat\rho\varphi)\right],
\end{equation}
one can get the following representation of the configuration integral in the thermodynamic limit $V\to \infty,~N_{i}\to\infty,~N_{i}/V\to c_{i},~N/V\to n$ in the form of a functional integral over the fluctuating electrostatic potential $\varphi(\bold{r})$ \cite{budkov2018nonlocal}:
\begin{equation}
\label{func_int1}
Q=\int\frac{\mathcal{D}\varphi}{C}\exp\left[-S[\varphi]\right],
\end{equation}
where the following auxiliary functional is introduced:
\begin{equation}
\nonumber
S[\varphi]=\frac{\beta}{2}(\varphi G_0^{-1}\varphi)-i\beta(\rho_{ext}\varphi)-n\int d\bold r\int d\bold r'\omega(\bold r-\bold{r}^{\prime})\left(e^{i\beta q(\varphi(\bold r)-\varphi(\bold{r}^{\prime}))}-1\right)
\end{equation}
\begin{equation}
-\sum\limits_{j=1}^{s}c_{j}\int d\bold r \left(e^{i\beta z_{j} e\varphi(\bold r)}-1\right)
\end{equation}
with the following shorthand notations
\begin{equation}
(f A f)=\int d\bold{r}\int d\bold{r}^{\prime}f(\bold{r})A(\bold{r},\bold{r}^{\prime})f(\bold{r}^{\prime}),~(fg)=\int d\bold{r}f(\bold{r})g(\bold{r}).
\end{equation}
The normalization constant $C=\int \mathcal{D}\varphi \exp\left[-\frac{\beta}{2}(\varphi G_0^{-1}\varphi)\right]$ is also introduced. The inverse Green function $G^{-1}_{0}(\bold{r},\bold{r}^{\prime})$ is determined by the following integral relation
\begin{equation}
\int d\bold{r}^{\prime\prime}G_{0}^{-1}(\bold{r},\bold{r}^{\prime\prime})G_{0}(\bold{r}^{\prime\prime}-\bold{r}^{\prime})=\delta(\bold{r}-\bold{r}^{\prime})
\end{equation}
that yields
\begin{equation}
G_{0}^{-1}(\bold{r},\bold{r}^{\prime})=-\frac{\varepsilon}{4\pi}\nabla^2\delta\left(\bold{r}-\bold{r}^{\prime}\right),
\end{equation}
where $\nabla^2=\Delta$ is the Laplace operator. It can be shown \cite{budkov2018nonlocal} that the obtained functional $S[\varphi]$ in the case of polar particles with a fixed distance between the charged centers, at its sufficiently small value, transforms into the well-known Poisson-Boltzmann-Langevin (PBL) functional \cite{abrashkin2007dipolar,coalson1996statistical,buyukdagli2014dipolar,budkov2015modified}
\begin{equation}
S_{PBL}[\varphi]=\frac{\beta}{2}(\varphi G_0^{-1}\varphi)-n\int d\bold r\left(\frac{\sin \beta p|\bold \nabla\varphi(\bold r)|}{\beta p|\bold \nabla\varphi(\bold r)|}-1\right)-i\beta(\rho_{ext}\varphi)-\int d\bold r \sum\limits_{j=1}^{s}c_{j}\left(e^{i\beta z_j e\varphi(\bold{r})}-1\right),
\end{equation}
where $p=ql$ is the dipole moment of the polar molecules.

\textbf{Mean-field approximation.}
In the mean field approximation \cite{naji2013perspective} the self-consistent field equation $\delta{S[\varphi_{MF}]}/\delta{\varphi(\bold{r})}=0$ for the electrostatic potential $\psi(\bold{r})=-i\varphi^{MF}(\bold{r})$ takes the following form
\begin{equation}
\label{scf_eq}
-\frac{\varepsilon}{4\pi}\Delta\psi(\bold r)=2 n q\int d\bold{r}^{\prime}\omega(\bold r-\bold{r}^{\prime})\sinh\frac{q(\psi(\bold r)-\psi(\bold r'))}{k_BT}+\sum\limits_{j=1}^{s}z_{j}ec_{j} e^{-\beta z_{j}e\psi(\bold{r})}+\rho_{ext}(\bold{r}).
\end{equation}
As is seen, equation (\ref{scf_eq}) is a generalization of the classical Poisson-Boltzmann equation \cite{naji2013perspective,levin2002electrostatic} for the self-consistent field potential for the case when the solution contains zwitterions. The electrostatic free energy in the mean-field approximation is
\begin{equation}
F_{el}^{(MF)}=k_BTS[i\psi].
\end{equation}

Then, solving self-consistent field equation (\ref{scf_eq}) within the approximation of the weak electrostatic interactions, i.e. when $q\psi(\bold r)/{k_BT}\ll1$, by the Fourier-transformation method, we arrive at the expression for the electrostatic potential of the test point-like ion with the charge density $\rho_{ext}(\bold{r})=q_0\delta(\bold{r})$ as follows
\begin{equation}
\label{pot_lin}
\psi(\bold r)=\frac{4\pi q_0}{\varepsilon}\int\frac{d\bold k}{(2\pi)^3}\frac{e^{i\bold k\bold r}}{k^2+\varkappa^2(\bold k)},
\end{equation}
where 
\begin{equation}
\label{screen_func_2}
\varkappa^2(\bold{k})=\kappa_{s}^2+\frac{8\pi n q^2}{\varepsilon k_{B}T}\left(1-\omega(\bold{k})\right)
\end{equation}
is the screening function \cite{brilliantov1993phase,khokhlov1982theory,borue1988statistical,budkov2018nonlocal,budkov2019Astatistical,budkov2019statistical} 
and
\begin{equation}
\omega(\bold k)=\int d\bold{r}e^{-i\bold{k}\bold{r}}\omega(\bold{r})
\end{equation}
is the characteristic function \cite{gnedenko2018theory} corresponding to the distribution function $\omega(\bold{r})$; $\kappa_s=(8\pi e^2 I/\varepsilon k_{B}T)^{1/2}$ is the inverse Debye length, attributed to the ionic species with the ionic strength $I=\sum_{j}z_{j}^2c_{j}/2$. For the model characteristic function \footnote{For the polar molecule with fixed dipole length $l$, it is necessary to use the characteristic function $\omega(\bold{k})=\sin{kl}/kl$, determining the following distribution function $\omega(r)=(4\pi l^2)^{-1}\delta(r-l)$ \cite{buyukdagli2014dipolar}. Using this characteristic function gives qualitatively the same results as for $\omega(\bold{k})=1/(1+k^2l^2/6)$, but does not allow us to obtain the analytical expression for the electrostatic free energy.}
\begin{equation}
\label{char_func_model}
\omega(\bold k)=\frac{1}{1+\frac{k^2l^2}{6}},
\end{equation}
determining the following distribution function
\begin{equation}
\label{soft_dip_char}
\omega(\bold r)=\frac{3}{2\pi l^2 r}\exp\left(-\frac{\sqrt{6}r}{l}\right),
\end{equation}
where $r=|\bold{r}|$, the following expression for the electrostatic potential can be obtained \cite{budkov2018nonlocal}
\begin{equation}
\label{potential_fin_2}
\psi(\bold{r})=\frac{q_0}{\varepsilon r}
\left(u(y_{d},y_{s})e^{-\kappa_1(y_{d},y_{s})r}+\left(1-u(y_{d},y_{s})\right)e^{-\kappa_2(y_{d},y_{s})r}\right),
\end{equation}
where I have introduced the following notations
\begin{equation}
\kappa_{1,2}(y_{d},y_{s})=\frac{\sqrt{3}}{l}\left(1+y_s+y_d \pm \sqrt{(1+y_s+y_d)^2-4y_s}\right)^{1/2},
\end{equation}
\begin{equation}
u(y_{d},y_{s})=\frac{y_s+y_d+\sqrt{(1+y_s+y_d)^2-4y_s}-1}{2\sqrt{(1+y_s+y_d)^2-4y_s}},
\end{equation}
and $y_{s}={\kappa_{s}^2l^2}/{6}$, $y_d={4\pi n q^2l^2}/{(3\varepsilon k_BT)}={l^2}/{6r_{D}^2}$; $r_{D}=\left(8\pi n q^2/\varepsilon k_{B}T\right)^{-1/2}$ is the Debye length associated with the charged centers of the zwitterions. Note that the potential (\ref{potential_fin_2}) monotonically decreases at long distances at all values of $y_{s}$ and $y_{d}$. It is interesting to analyze the limiting regimes, following from eq. (\ref{potential_fin_2}). 
For the salt-free solution ($y_s =0$) we get:
\begin{equation}
\label{potential_fin}
\psi(\bold{r})=\frac{q_0}{\varepsilon_{l}(r) r}
\end{equation}
with the local dielectric permittivity being
\begin{equation}
\label{eps}
\varepsilon_{l}(r)=\frac{\varepsilon(1+y_d)}{1+y_d\exp\left(-\frac{r}{l_s}\right)}
\end{equation}
and $l_s=l/\sqrt{6(1+y_d)}$ is the new length scale which determines the radius of the sphere with the point-like test ion in its center, within which the dielectric permittivity is smaller than its bulk value, determined by the relation
\begin{equation}
\label{eps_bulk}
\varepsilon_b=\varepsilon(1+y_d)=\varepsilon+\frac{4\pi q^2l^2}{3k_BT}n.
\end{equation}
Eq. (\ref{eps_bulk}) for the bulk dielectric permittivity is the same as the one obtained in paper \cite{ramshaw1980debye} by means of the integral equations theory within the mean spherical approximation for the rigid dipoles. Thus, the length $l_{s}$ can be interpreted as the effective solvation radius of the point-like charge surrounded by zwitterions, within the linear approximation. Note that at $y_d\gg 1$ eq. (\ref{potential_fin}) transforms into the standard Debye-Hueckel (DH) potential
\begin{equation}
\psi(\bold{r})=\frac{q_{0}}{\varepsilon r}\exp\left[-{r}/{r_{D}}\right].
\end{equation}
Thus, when the dipole length $l$ is much bigger than the Debye length $r_{D}$, the charged centers of zwitterions manifest themselves as unbound ions, participating in the screening of the test point-like charge $q_0$.  
\begin{figure}[h!]
\center{\includegraphics[width=0.6\linewidth]{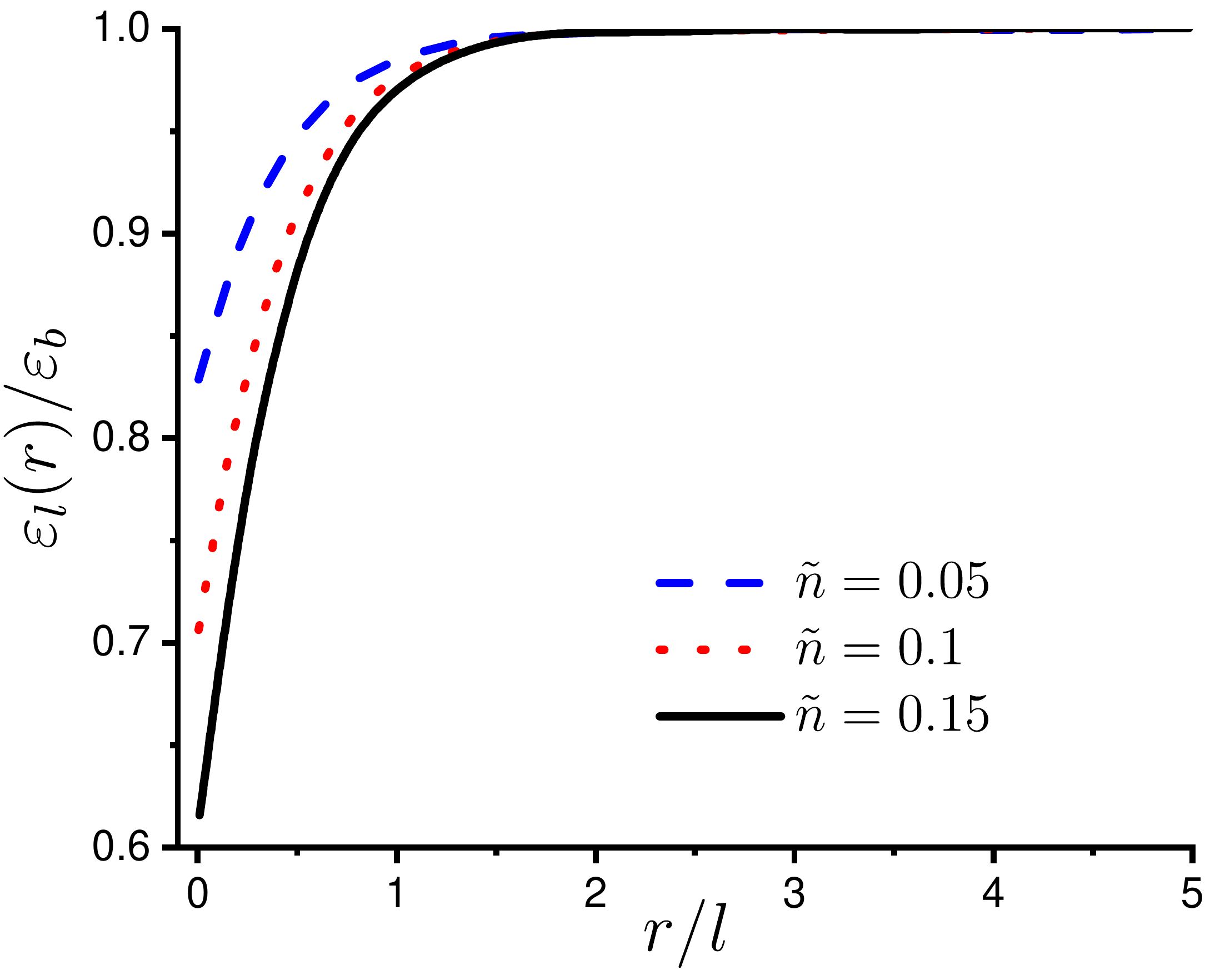}}
\caption{Local dielectric permittivity around the point-like test ion as a function of distance, expressed in dipole length $l$ units, for different reduced concentrations of zwitterions $\tilde{n}=nl^3$.}
\label{potential}
\end{figure}
In the absence of the zwitterions ($y_{d}=0$) we obtain the standard Debye-Hueckel potential \cite{landau2013course}:
\begin{equation}
\psi(\bold{r})=\frac{q_0}{\varepsilon r}e^{-\kappa_{s}r}.
\end{equation}

Then, after simple algebraic transformations, one can show \cite{budkov2018nonlocal} that the mean-field electrostatic free energy within the linear approximation is:
\begin{equation}
\label{el_free_en_lin_1}
F_{el}^{(MF)}=\frac{1}{2}\int d\bold r\rho_{ext}(\bold r)\psi(\bold r).
\end{equation}
To obtain an expression for the solvation energy of the test ion, one must subtract its self-energy from the electrostatic free energy, i.e.
\begin{equation}
\nonumber
\Delta F_{el}=\mu_{el}^{(ion)}=\frac{1}{2}\int d\bold r\rho_{ext}(\bold r)\left(\psi(\bold r)-\psi_{ext}(\bold r)\right)
\end{equation}
\begin{equation}
\label{solv_1}
=-\frac{q_0^2}{2\varepsilon}\left(u(y_{d},y_{s})(\kappa_1(y_{d},y_{s})-\kappa_2(y_{d},y_{s}))+\kappa_2(y_{d},y_{s})\right).
\end{equation}
I would like to note that the solvation energy of the point-like charge is equal to its excess chemical potential (see below). In the absence of the salt ions ($y_s=0$), the expression for the solvation energy (\ref{solv_1}) of the test charge $q_0$ takes the form \cite{budkov2018nonlocal}
\begin{equation}
\label{solv_2}
\mu_{el}^{(ion)}=-\frac{q_0^2}{2\varepsilon R_{ef}},
\end{equation}
where
\begin{equation}
R_{ef}=\frac{l}{\sqrt{6}} \frac{\sqrt{1+y_d}}{y_d}
\end{equation}
is the effective radius of the point-like charge in the environment of the polar molecules. It is interesting to note that in contrast to the local theory with point-like dipoles, within this nonlocal theory, the electrostatic free energy of the point-like test ion has a finite value. Note that at $y_d\ll 1$ (or at $l\to 0$) the solvation energy (\ref{solv_2}) tends to zero linearly with $l$, i.e.
\begin{equation}
\mu_{el}^{(ion)}\simeq -\frac{2\pi \sqrt{6}}{3}\frac{q^2 q_0^2n}{\varepsilon^2 k_{B}T}l.
\end{equation}
In the opposite case, $y_{d}\gg 1$, when the charged sites of zwitterions behave as unbound ions the solvation energy results in
\begin{equation}
\mu_{el}^{(ion)}\simeq -\frac{q_0^2}{2\varepsilon r_{D}},
\end{equation}
which is the well known expression for excess chemical potential of ion in dilute electrolyte solution within the DH theory \cite{landau2013course}. I would like to note that eq. (\ref{solv_2}) describes the solvation energy when we can neglect the linear size of solvated ion, i.e. when $l\gg a$, where $a$ is the ionic radius. Indeed, within the present theory, the point-like dipole limit ($l\to 0$) with keeping $ql$ constant gives infinite solvation energy of point-like test ion. Such a divergence of solvation energy is related to the fact that I do not take into account excluded volume of test ion. Accounting for the excluded volume of solvated ion will allow us to obtain a finite value of solvation energy even in point-like dipole limit \cite{levin2002electrostatic,levin1999happened,adar2018dielectric}.

\textbf{Random phase approximation.}
The simplest method to estimate the electrostatic free energy of salt zwitterionic solution is the random phase approximation (RPA) \cite{naji2013perspective,budkov2018nonlocal}. Expanding the functional $S[\varphi]$ in eq. (\ref{func_int1}) into a power functional series in the neighbourhood of the mean-field configuration $\varphi_{MF}(\bold{r})=i\psi(\bold{r})$ and truncating it by the quadratic term on the functional variable $\Lambda(\bold{r})=\varphi(\bold{r}) - i\psi(\bold{r})$, I obtain:
\begin{equation}
Q\approx \exp\left[-S[i\psi]\right]\int\frac{\mathcal{D}\Lambda}{C}\exp\left[-\frac{\beta}{2}\left(\Lambda \mathcal{G}^{-1}\Lambda\right)\right],
\end{equation}
where
\begin{equation}
\nonumber
\mathcal{G}^{-1}(\bold r,\bold r'|\psi)=k_{B}T\frac{\delta^2 S[i\psi]}{\delta \varphi(\bold{r})\delta \varphi(\bold{r}')}=G_0^{-1}(\bold r,\bold r')+\beta e^2\sum\limits_{j=1}^{s}z_{j}^2e^{-\beta z_{j}e\psi(\bold{r})}\delta(\bold r-\bold r')
\end{equation}
\begin{equation}
\label{dip_green}
+\frac{2n q^2}{k_BT}\int d\bold{r}''\omega(\bold{r}-\bold{r}'')\cosh\left(\frac{q\left(\psi(\bold{r})-\psi(\bold{r}'')\right)}{k_{B}T}\right)(\delta(\bold r-\bold r')-\delta(\bold r''-\bold r'))
\end{equation}
is the renormalized inverse Green function depending on the mean-field electrostatic potential $\psi(\bold{r})$, satisfying eq. (\ref{scf_eq}). Thus, when the Gaussian functional integral is evaluated by the standard methods \cite{zinn1996quantum}, the following general relation for the configuration integral within the RPA can be obtained:
\begin{equation}
\label{RPA_gen}
Q\approx \exp\left[-S[i\psi]+\frac{1}{2}tr\left(\ln \mathcal{G} -\ln G_{0}\right)\right],
\end{equation}
where the symbol $tr(...)$ means the trace of the integral operator \cite{budkov2019statistical,budkov2019Astatistical}, i.e.
$tr(A)=\int d\bold{r} A(\bold{r},\bold{r})$, where $A(\bold{r},\bold{r}^{\prime})$ is the kernel of integral operator $A$. In the absence of the external charges (i.e., at $\rho_{ext}(\bold{r})=0$), the electrostatic potential $\psi(\bold{r})=0$, and the mean-field contribution to the electrostatic free energy $F_{el}^{(MF)}=k_{B}T S[0]=0$. Thus, in this case, the electrostatic free energy is determined by the thermal fluctuations of the electrostatic potential near its zero value and takes the following form (see Appendix I):
\begin{equation}
\label{el_free_en_lin_2}
F_{el}=\frac{Vk_BT}{2}\int\frac{d\bold k}{(2\pi)^3}\left(\ln\left(1+\frac{\varkappa^2(\bold k)}{k^2}\right)-\frac{\varkappa^2(\bold k)}{k^2}\right),
\end{equation}
where the screening function $\varkappa^2(\bold{k})$ is determined by eq. (\ref{screen_func_2}). The excess osmotic pressure caused by the electrostatic interactions which can be obtained from the standard thermodynamic relation \cite{landau2013course} $\Pi_{el}=-\left(\partial{F_{el}}/\partial{V}\right)_T$, takes the form
\begin{equation}
\label{osm_press}
\Pi_{el}=\frac{k_{B}T}{2}\int\frac{d\bold{k}}{(2\pi)^3}\left(\frac{\varkappa^2(\bold{k})}{k^2+\varkappa^2(\bold{k})}-\ln\left(1+\frac{\varkappa^2(\bold{k})}{k^2}\right)\right).
\end{equation}
For the distribution function, defined by eq. (\ref{char_func_model}), in the absence of the ions in the solution ($I=0$), the integral in eq. (\ref{el_free_en_lin_2}) can be calculated analytically \cite{budkov2018nonlocal}:
\begin{equation}
\label{RPA_pure_dip}
F_{el}=-\frac{Vk_BT}{l^3}\sigma(y_{d}),
\end{equation}
where the auxiliary dimensionless function  $\sigma(y_d)=\sqrt{6}(2(1+y_{d})^{3/2}-2-3y_{d})/4\pi$ is introduced. As is seen, accounting for the details of the internal electric structure of the zwitterions allows us to obtain a finite value of the excess free energy without an artificial cut-off. The excess osmotic pressure for the salt-free solution takes the following analytical form
\begin{equation}
\label{RPA_pure_dip_2}
\Pi_{el}=-\frac{k_{B}T}{l^3}\sigma_{1}(y_d)
\end{equation}
with the auxiliary dimensionless function $\sigma_1(y_d)=\sqrt{6}(3y_d(1+y_{d})^{1/2}/2-(1+y_{d})^{3/2}+1)/2\pi$.
The excess free energy and osmotic pressure of the salt-free solution of the zwitterions can be analysed in two limiting regimes, resulting from Eqs. (\ref{RPA_pure_dip}-\ref{RPA_pure_dip_2}), namely:
\begin{equation}
\frac{F_{el}}{Vk_BT}=
\begin{cases}
-\frac{\sqrt{6}\pi q^4 l}{3 \left(\varepsilon k_{B}T\right)^2} n^2,~y_d\ll 1\,\\
-\frac{1}{12\pi r_{D}^3},~y_d\gg 1,
\end{cases}
\end{equation}
and
\begin{equation}
\frac{\Pi_{el}}{k_BT}=
\begin{cases}
-\frac{\sqrt{6}\pi q^4 l}{3 \left(\varepsilon k_{B}T\right)^2} n^2,~y_d\ll 1\,\\
-\frac{1}{24\pi r_{D}^3},~y_d\gg 1.
\end{cases}
\end{equation}
In the first regime, the $"$gas$"$ of the polar molecules, interacting via the simple dipole-dipole pair potential, is realized, whereas in the second regime, the charged groups of the zwitterions can be considered as unbound ions, and the electrostatic free energy and excess osmotic pressure are described by the DH limiting law. In other words, when the dipole length is much bigger than the Debye length, the charged groups do not $"$feel$"$ any more that they are parts of the zwitterions and, thereby, behave as freely moving ions.

Then, the limiting regimes of the behavior of the excess free energy and osmotic pressure in the presence of the salt ions at $y_{d}\ll 1$ and $y_{d}\gg 1$ can be also analysed. At $y_{d}\ll 1$, the results are
\begin{equation}
\label{free}
\frac{F_{el}}{Vk_{B}T}=-\frac{\kappa_{s}^3}{12\pi}-\frac{3\sqrt{6}}{2\pi l^3}\frac{y_{s}}{1+\sqrt{y_s}}y_d-\frac{3\sqrt{6}}{16\pi l^3}\frac{1+3\sqrt{y_s}+y_s}{(1+\sqrt{y_s})^3}y_{d}^2+O(y_{d}^3)
\end{equation}
and
\begin{equation}
\label{press}
\frac{\Pi_{el}}{k_{B}T}=-\frac{\kappa_{s}^3}{24\pi}-\frac{3\sqrt{6}}{4\pi l^3}\frac{y_{s}(2+\sqrt{y_s})}{(1+\sqrt{y_s})^2}y_d-\frac{3\sqrt{6}}{16\pi l^3}\frac{2+8\sqrt{y_s}+4y_s+y_s^{3/2}}{2(1+\sqrt{y_s})^4}y_{d}^2+O(y_{d}^3),
\end{equation}
where the first terms on the right hand side of Eqs. (\ref{free}-\ref{press}) describe the contribution of the Coulomb interaction of the ionic species to the excess free energy and osmotic pressure within the DH approximation. The second and third terms describe the contributions of the ion-dipole and dipole-dipole pair correlations, respectively. In the opposite regime, when $y_d\gg 1$, we arrive at the DH limiting law:
\begin{equation}
\frac{F_{el}}{Vk_{B}T}\simeq-\frac{\kappa^3}{12\pi},~\frac{\Pi_{el}}{k_{B}T}\simeq-\frac{\kappa^3}{24\pi},
\end{equation}
where $\kappa=\left(8\pi(q^2n +e^2I)/\varepsilon k_{B}T\right)^{1/2}$ is the inverse Debye length. It is worth noting that in this case the charged groups of the zwitterionic molecules behave as unbound ions participating in the electrostatic screening along with the salt ions.

Eq. (\ref{el_free_en_lin_2})  makes it possible to calculate the expression for the excess chemical potential (solvation energy) of the point-like ion with the charge $z_i e$. Using the standard thermodynamic relation $\mu_{el,j}^{(ion)}=\partial{F_{el}}/\partial{N_j}$, we arrive at
\begin{equation}
\label{mu_el}
\mu_{el,j}^{(ion)}=-\frac{z_{j}^2e^2}{\pi \varepsilon}\int\limits_{0}^{\infty}dk \frac{\varkappa^{2}(\bold{k})}{k^2+\varkappa^{2}(\bold{k})}.
\end{equation}
Further, using Eqs. (\ref{screen_func_2}) and (\ref{char_func_model}) and calculating the integral in eq. (\ref{mu_el}), we arrive at eq. (\ref{solv_1}) at $q_0=z_{i}e$.

\textbf{Multicomponent zwitterionic salt solutions.}
The above theory can be easily generalized for the case of multicomponent salt solutions of zwitterions. Let me consider a salt solution with $p$ kinds of zwitterions with the total numbers $N_{k}$ ($k=1,2,...,p$). I also assume that a certain distribution function $\omega_{k}(\bold{r})$ ($k=1,2,...,p$) of distance between the charged centers is associated with each type of zwitterion. Performing the same mathematical manipulations as described above, I can recast the configuration integral of the solution in the thermodynamic limit as follows
\begin{equation}
Q=\int\frac{\mathcal{D}\varphi}{C}\exp\left[-S[\varphi]\right],
\end{equation}
where the functional $S[\varphi]$ has the following form
\begin{equation}
\nonumber
S[\varphi]=\frac{\beta}{2}(\varphi G_0^{-1}\varphi)-i\beta(\rho_{ext}\varphi)-W[\varphi],
\end{equation}
where
\begin{equation}
W[\varphi]=\int d\bold r\int d\bold{r}^{\prime}\sum\limits_{k=1}^{p}n_{k}\omega_{k}(\bold r-\bold r')(e^{i\beta q_{k}(\varphi(\bold r)-\varphi(\bold r'))}-1)+\int d\bold r\sum\limits_{j=1}^{s}c_{j}(e^{i\beta z_{j}e\varphi(\bold r)}-1),
\end{equation}
$n_{k}=N_{k}/V$ ($k=1,2,...,p$) are the average concentrations of the zwitterions of the $k$-th type, and $q_{k}$ are the charges of their charged centers. The self-consistent field equation for this case has the following form
\begin{equation}
\label{scf_eq_2}
\Delta\psi(\bold r)=-\frac{4\pi}{\varepsilon}\left(\bar{\rho}_{zw}(\bold{r})+\bar{\rho}_{i}(\bold{r})+\rho_{ext}(\bold{r})\right),
\end{equation}
where the average bound charge density of the zwitterions
\begin{equation}
\bar{\rho}_{zw}(\bold{r})=2\sum\limits_{k=1}^{p}n_{k}q_{k} \int d\bold{r}^{\prime} \omega_{k}(\bold r-\bold{r}^{\prime})\sinh\frac{q_{k}(\psi(\bold r)-\psi(\bold r'))}{k_BT}
\end{equation}
and the average charge density of the ions
\begin{equation}
\label{av_ch_dens}
\bar{\rho}_{i}(\bold{r})=\sum\limits_{j=1}^{s}z_{j}ec_{j} e^{-\beta z_{j}e\psi(\bold{r})}
\end{equation}
have been introduced. The excess free energy and osmotic pressure can be determined by Eqs. (\ref{el_free_en_lin_2}) and (\ref{osm_press}) with the screening function being
\begin{equation}
\varkappa^2(\bold{k})=\kappa_s^2+\frac{8\pi}{\varepsilon k_{B}T}\sum\limits_{k=1}^{p}q_{k}^2n_{k}\left(1-\omega_{k}(\bold{k})\right),
\end{equation}
where $\omega_{k}(\bold{k})$ are the characteristic functions. Thus, using the latter results, I obtain the following general relation for the excess chemical potential of the zwitterionic molecules of the $j$-th kind
\begin{equation}
\label{mu_el_2}
\mu_{el,j}^{(zw)}=-\frac{2q_{j}^2}{\pi \varepsilon}\int\limits_{0}^{\infty}dk(1-\omega_{j}(\bold{k}))\frac{\varkappa^{2}(\bold{k})}{k^2+\varkappa^{2}(\bold{k})}.
\end{equation}

It is instructive to apply eq. (\ref{mu_el_2}) to the case when a test zwitterion with charges $\pm q_0$ and a distribution function $\omega_{0}(r)$ is dissolved in the salt-free solution of the zwitterions. For this purpose, it is necessary to consider a solution with two types of zwitterions with charges $\pm q_{0}$ and $\pm q$ and distribution functions $\omega_{0}(\bold{r})$ and $\omega(\bold{r})$. Assuming that the concentration $n_0$ of the zwitterions with charges $\pm q_0$ tends to zero, I obtain the following expression for the excess chemical potential in the limit of an infinitely dilute solution
\begin{equation}
\label{mu_el_3}
\mu_{el}^{(zw)}=-\frac{16q_{0}^2q^2n}{\varepsilon^2 k_{B}T}\int\limits_{0}^{\infty}dk\frac{(1-\omega_{0}(\bold{k}))(1-\omega(\bold{k}))}{k^2+\varkappa^{2}(\bold{k})},
\end{equation}
where $\varkappa^2(\bold{k})=8\pi q^2n\left(1-\omega(\bold{k})\right)/(k_{B}T\varepsilon)$ and $n$ is the concentration of zwitterions with charges $\pm q$. For the characteristic functions, determined by the expressions $\omega_{0}(\bold{k})=(1+k^2l_0^2/6)^{-1}$ and $\omega(\bold{k})=(1+k^2l^2/6)^{-1}$, I arrive at the following analytical expression
\begin{equation}
\label{mu_el_4}
\mu_{el}^{(zw)}=-\frac{\sqrt{6}q_0^2}{\varepsilon l}\frac{\gamma y_d}{1+\gamma\sqrt{1+y_d}},
\end{equation}
where $\gamma=l_0/l$. It is interesting to note that in the regime $\gamma \gg 1$ (a very long dipole of the solvated zwitterion), the zwitterion solvation energy is twice as high as the solvation energy of the point-like test ion with the charge $q_0$ (see eq. (\ref{solv_2})). The latter can be explained by the fact that in this case the distance between the two ionic groups is very large, so that they behave as two independent ions.

\textbf{Zwitterions with a hard core.} 
Above, I discussed rather dilute solutions of zwitterions, for which it is safe to neglect the short-range interactions of the molecules that are related to their excluded volume and take into account only the long-range electrostatic interactions. However, at a sufficiently high concentration of the solution the short-range interactions also become important. Let me discuss how the excluded volume interactions between zwitterions can be taken into account. In this section I will consider a salt-free solution of zwitterions, describing them within the model of dipolar hard spheres. I would like to note that I have recently formulated \cite{budkov2019statistical} the model of the salt-free solution of zwitterions with a soft core described by the repulsive Gaussian potential. 

\begin{figure}[h!]
\center{\includegraphics[width=0.55\linewidth]{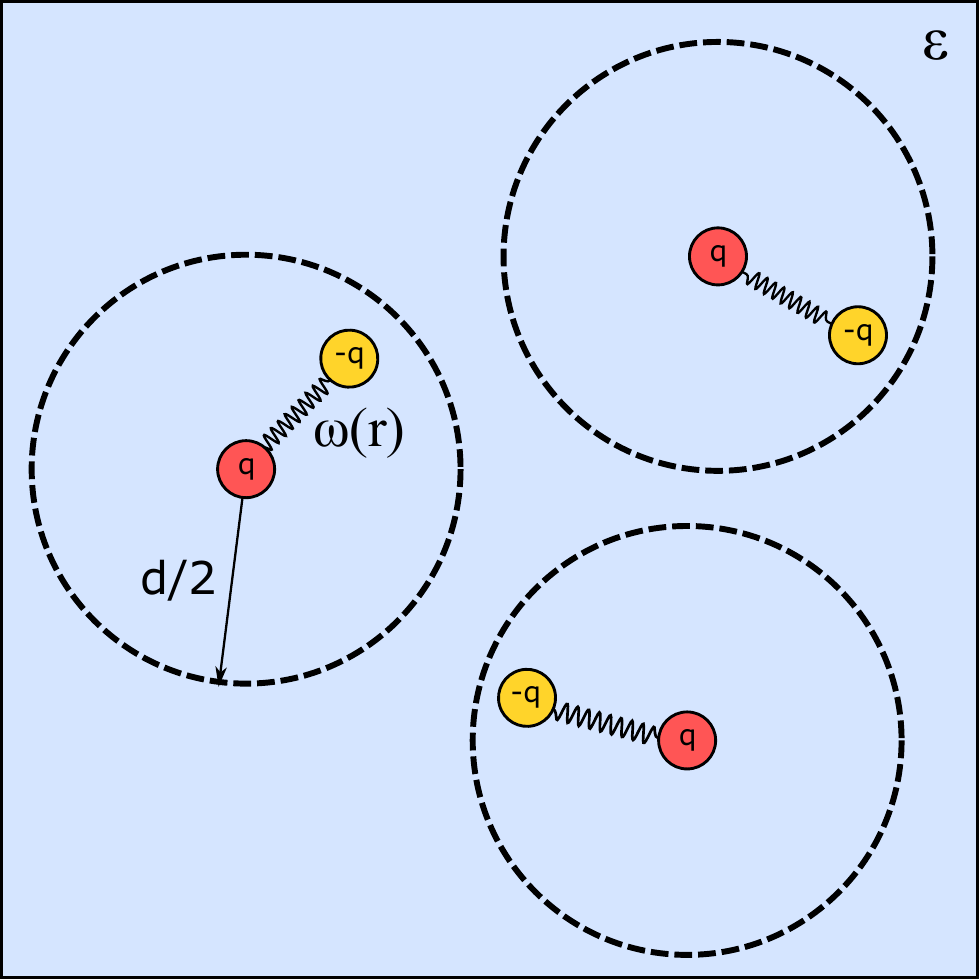}}
\caption{Schematic representation of a salt-free solution of zwitterions with a hard core.}
\label{potential}
\end{figure}

A point-like charge $q$ is placed at the center of each hard sphere with a diameter $d$ and an oppositely charged site $-q$ is grafted to the center of the sphere. Let me also assume that the charged sites $\pm q$ are separated by a fluctuating distance described by a probability distribution function $\omega(\bold{r})$. Let me assume also that the charged site $-q$ is a phantom one, so that it can penetrate inside the hard cores of molecules. The screening function of such a system can be written in the following form (for details, see Appendix II)
\begin{equation}
\label{varkappa2}
\varkappa^2(\bold{k})=\frac{8\pi n q^2}{\varepsilon k_{B}T}\left(1-\omega(\bold{k})\right)\left(1+\frac{1}{2}\left(S(\bold{k})-1\right)\left(1-\omega(\bold{k})\right)\right),
\end{equation}
where 
\begin{equation}
\label{S}
S(\bold{k})=1+nh(\bold{k})
\end{equation}
is the structure factor of the hard spheres system expressed via the Fourier-image $h(\bold{k})$ of the pair correlation function \cite{hansen1990theory}. The electrostatic free energy is determined by eq. (\ref{el_free_en_lin_2}) with screening function (\ref{varkappa2}). We would like to note that the structure factor (\ref{S}) of hard spheres system does not determine the total structure factor of zwitterions. The total structure factor, taking into account dipole-dipole interactions, can behave different from the structure factor of hard spheres. Further, using the $"$mean-field$"$ approximation $S(\bold{k})\approx S(0)=nk_{B}T\chi_{T}=z_{0}$, where $\chi_{T}$ is the isothermal compressibility of the hard spheres system, and using the characteristic function $\omega(\bold{k})=(1+k^2l^2/6)^{-1}$, one can obtain an analytical expression for the electrostatic free energy of the dipolar hard spheres \cite{budkov2020nonlocal}. I do not provide a rather cumbersome expression for electrostatic free energy here and write only the limiting regimes following from it
\begin{equation}
\label{limit_reg}
\frac{F_{el}}{Vk_BT}=
\begin{cases}
-\frac{\sqrt{6}\pi q^4 l}{3 \left(\varepsilon k_{B}T\right)^2} n^2\left(1-\frac{3}{4}\alpha+\frac{5\alpha^2}{32}\right),~y_d\ll 1\,\\
-\frac{1}{12\pi r_{D}^3}\left(1-\frac{3}{2}\alpha-3(1-e^{\alpha/4})\right),~y_d\gg 1,
\end{cases}
\end{equation}
where $\alpha=1-z_0$. In the Percus-Yevick \cite{hansen1990theory,Gray} approximation I get
\begin{equation}
\alpha=\frac{\eta(4-\eta)(2+\eta^2)}{(1+2\eta)^2},
\end{equation}
where $\eta=\pi d^3 n/6$ is the packing fraction of the hard spheres \cite{hansen1990theory}.
Using limiting regimes (\ref{limit_reg}), one can construct the approximate expression \cite{budkov2020nonlocal} which quite well reproduces the electrostatic free energy for all values of $y_d$ and $0<\alpha<1$:
\begin{equation}
\label{Fel_dip_hs}
F_{el}\approx-\frac{Vk_{B}T}{l^3}\left(1-\frac{3}{4}\alpha\right)\sigma(y_d),
\end{equation}
As one can see from the comparison of eqs. (\ref{Fel_dip_hs}) and (\ref{RPA_pure_dip}), accounting for the excluded volume of the zwitterions reduces the electrostatic contribution to the total free energy. The latter can be explained by the fact that the hard core of the molecules makes their overlapping impossible, thus, decreasing the electrostatic interactions.

The adopted $"$mean-field$"$ approximation $S(\bold{k})\approx S(0)$ would be valid when the contribution of $k \ll 2 \pi/d$ makes dominant contribution in integrals over $\bold{k}$, which means that the interaction between molecules at distances much larger than $d$ is important. However, for the systems of high molecular concentration this approximation is not valid. In this case, it is necessary to integrate numerically over $\bold{k}$, using analytical expression for structure factor of hard sphere systems within the Percus-Yiewick approximation \cite{santos2016exact}. Moreover, we would like to note that RPA is valid at sufficiently small electrostatic interactions between charged particles \cite{naji2013perspective}. Only in this case we can use the hard spheres fluid as a reference system, considering the electrostatic contribution as perturbation.

\section{Solutions of multipolar molecules}
In the previous section, I considered the general statistical field theory of salt solutions of zwitterions, which might be used for describing the thermodynamic properties of real ion-molecular solutions, such as proteins and betaines, whose molecules possess large dipole moments. However, there is a wide class of macromolecules, acquiring a complex electric structure in solvent media, which cannot be reduced to two oppositely charge centers -- a dipolar structure. As it was already mentioned in the introductory section, the most natural generalization of the dipolar structure is the structure of a $"$star$"$, i.e., when the $"$central$"$ charge $q$ is placed at the center of the molecule, surrounded by $"$peripheral$"$ charges $q_{\alpha}$ ($\alpha=1,...,m$). The positions of the peripheral charges relative to the central charge are fixed or fluctuating ones. Each molecule is assumed to be electrically neutral, so that the neutrality condition $q+\sum_{\alpha}q_{\alpha}=0$ is fulfilled. As I have already mentioned in the Introduction, such multipolar configurations could be realized for macromolecular systems, such as polyelectrolyte brushes and stars, complex colloids, and metal-organic complexes. Despite the fact that there is a wide range of possible applications, up to now, there have been no attempts to develop the statistical theories of salt solutions of multipolar molecules \cite{budkov2019Astatistical}. In this section, I would like to consider the FT model of salt solutions of neutral molecules, possessing a multipolar electric structure. As in the case of the above described solutions of zwitterionic molecules, I assume that the ions with total numbers $N_{i}$ and valencies $z_i$ ($i=1,...,s$) are dissolved in a solvent medium along with multipolar molecules. As in the previous section, the global neutrality condition $\sum_{i}z_{i}N_{i}=0$ is fulfilled and the solvent is supposed to be a continuous dielectric medium with a constant permittivity $\varepsilon$. I assume that each peripheral charge $q_{\alpha}$ is separated from the central charge $q$ by the fluctuating distance $\bold{\xi}_{\alpha}$, described by the distribution function $\omega_{\alpha}(\bold{\xi}_{\alpha})$ ($\alpha=1,..,m$). I also assume that the solution is rather dilute, the excluded volume interactions between all the species can be neglected. The effects of excluded volume interactions will be discussed at the end of this section.

\begin{figure}[h!]
\center{\includegraphics[width=0.55\linewidth]{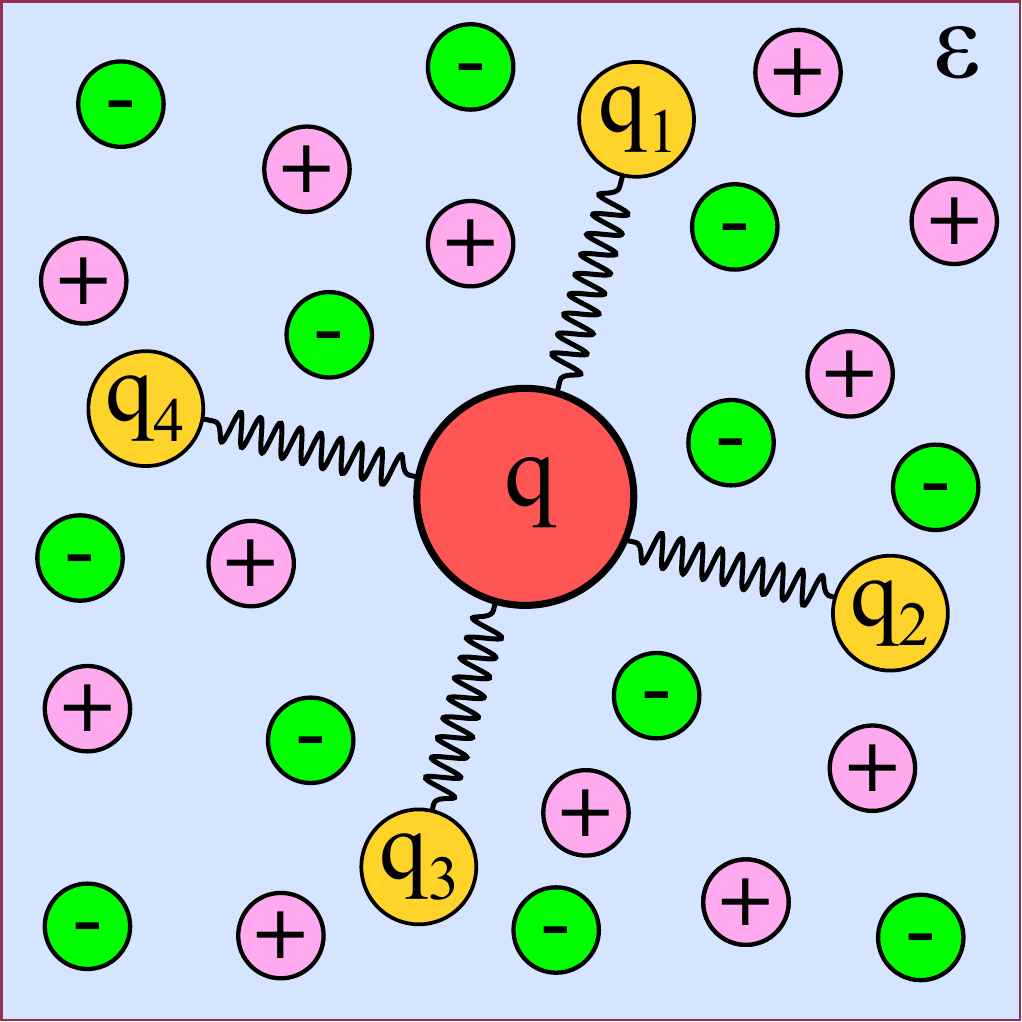}}
\caption{Schematic representation of the multipolar molecule surrounded by salt ions in a dielectric solvent medium.}
\label{potential}
\end{figure}

Taking into account all the model assumptions, one can write the configuration integral of the system in the form
\begin{equation}
Q=\int d\Gamma_s\int d\Gamma_{h}\exp\left[-\beta H_{int}\right],
\end{equation}
where
\begin{equation}
\label{dGamma_h}
\int d\Gamma_h(\cdot)=\int\dots\int\prod\limits_{j=1}^{N} d\Gamma_{j}(\cdot)
\end{equation}
is the integration measure over the configurations $\Gamma_{j}$ of the $"$hairy$"$ multipolar molecules
\begin{equation}
\label{dGamma_j}
\int d\Gamma_{j}(\cdot)=
\int\frac{d\bold{R}_j}{V}\int\dots\int\prod\limits_{\alpha=1}^{m}d\bold{\xi}_{j}^{(\alpha)}\omega_{\alpha}\left(\xi_{j}^{(\alpha)}\right)(\cdot),
\end{equation}
where $\bold{\xi}_{j}^{(\alpha)}$ are the displacement vectors of the peripheral charges relative to the positions $\bold{R}_{j}$ of the central charges. The measure of integration over the positions of the salt ions is determined by eq. (\ref{salt_int_meas}).
The total potential energy of the electrostatic interactions is
\begin{equation}
\label{hamilt_el}
H_{int}=\frac{1}{2}\int d\bold{r}\int d\bold{r}'\hat\rho(\bold{r})G_0(\bold{r}-\bold{r}')\hat\rho(\bold{r}')=\frac{1}{2}\left(\hat\rho G_0 \hat\rho\right),
\end{equation}
where
\begin{equation}
\hat\rho(\bold r)=\hat\rho_{h}(\bold r)+\hat\rho_{i}(\bold r)+\rho_{ext}(\bold r)
\end{equation}
is the total charge density of the system, where
\begin{equation}
\label{rho_h}
\hat\rho_h(\bold r)=\sum\limits_{\alpha=1}^{m}q_{\alpha}\sum\limits_{j=1}^{N}
\left(\delta\left(\bold{r}-\bold{R}_{j}-\bold{\xi}_{j}^{(\alpha)}\right)-\delta\left(\bold{r}-\bold{R}_{j}\right)\right)
\end{equation}
is the local charge density of the multipolar molecules and $\hat{\rho}_{i}(\bold{r})$ is determined by eq. (\ref{ch_dens_ion}).
Using HS transformation (\ref{HS}), I arrive in the thermodynamic limit at
\begin{equation}
\label{func_int}
Q=\int\frac{\mathcal{D}\varphi}{C}\exp\left[-S[\varphi]\right],
\end{equation}
where
\begin{equation}
S[\varphi]=\frac{\beta}{2}(\varphi G_0^{-1}\varphi)-i\beta(\rho_{ext}\varphi)-W[\varphi]
\end{equation}
with 
\begin{equation}
W[\varphi]=n\int d\bold R \left<e^{i\beta\sum\limits_{\alpha=1}^{m}q_{\alpha} (\varphi(\bold{R}+\bold{\xi}_{\alpha})-\varphi(\bold{R}))}-1\right>_{\xi}+\sum\limits_{j=1}^{s}c_{j}\int d\bold r (e^{i\beta z_{j} e\varphi(\bold r)}-1).
\end{equation}
Note that in the latter equation I introduced notation for averaging over the configurations of the peripheral charges
\begin{equation}
\left<(\cdot)\right>_{\xi}=\int d\bold{\xi}_{1}\dots\int d\bold{\xi}_{m}\prod\limits_{\alpha=1}^{m}\omega_{\alpha}(\bold{\xi}_{\alpha})(\cdot).
\end{equation}

\textbf{Oscillations of the electrostatic potential.}
The self-consistent field equation for the salt solution of the multipolar molecules takes the form \cite{budkov2019Astatistical}
\begin{equation}
\label{SC_eq}
\Delta\psi(\bold{r})=-\frac{4\pi}{\varepsilon}
\left(\rho_{ext}(\bold{r})+\bar{\rho}_{i}(\bold{r})+\bar{\rho}_{h}(\bold{r})\right),
\end{equation}
where $\bar{\rho}_{i}(\bold{r})$ is the average charge density of the ions, determined by eq. (\ref{av_ch_dens}) and
\begin{equation}
\label{rho_c}
\bar{\rho}_{h}(\bold{r})=n \sum\limits_{\gamma=1}^{m}q_{\gamma}
\left<e^{-\beta\sum\limits_{\alpha=1}^{m}q_{\alpha}\left(\psi(\bold{r}+\bold{\xi}_{\alpha}-\bold\xi_{\gamma})-\psi(\bold{r}-\bold\xi_{\gamma})\right)}-e^{-\beta\sum\limits_{\alpha=1}^{m}q_{\alpha}\left(\psi(\bold{r}+\bold{\xi}_{\alpha})-\psi(\bold{r})\right)}\right>_{\xi}
\end{equation}
is the average charge density of the multipolar hairy molecules.

In the linear approximation, eq. (\ref{SC_eq}) yields the following expression for the electrostatic potential of the external charges
\begin{equation}
\label{psi}
\psi(\bold{r})=\frac{4\pi}{\varepsilon}\int\frac{d\bold{k}}{(2\pi)^3}\frac{ \tilde{\rho}_{ext}(\bold{k})e^{i\bold{k}\bold{r}}}{k^2+\varkappa^2(\bold{k})},
\end{equation}
with the following screening function
\begin{equation}
\label{screen_func_3}
\varkappa^2(\bold{k})=\kappa_{s}^2+\frac{4\pi n}{\varepsilon k_{B}T}\left(|q(\bold{k})|^2+\sum\limits_{\alpha=1}^{m}q_{\alpha}^2(1-|\omega_{\alpha}(\bold{k})|^2)\right),
\end{equation}
and the Fourier-image of the density of the external charges $\tilde{\rho}_{ext}(\bold{k})$.
I have introduced the characteristic functions 
\begin{equation}
\omega_{\alpha}(\bold{k})=\int d\bold{r} \omega_{\alpha}(\bold{r}) e^{-i\bold{k}\bold{r}}
\end{equation} 
and the auxiliary function
\begin{equation}
\label{q}
q(\bold{k})=\sum_{\alpha=1}^{m}q_{\alpha}(1-\omega_{\alpha}(\bold{k})).
\end{equation} 
Note that for $m=1$, screening function (\ref{screen_func_3}) transforms into the above written screening function (\ref{screen_func_2}) for the zwitterionic molecules. Note also that within the linear approximation (\ref{psi}) the electrostatic correlations between peripheral charges within one multipolar molecule are not taken into account. The latter can be justified by sufficiently small electrostatic interactions which unable to disturb strongly the configuration of multipolar molecules.

Using eq. (\ref{psi}), one can analyse the behavior of the potential of the point-like test ion $q_{0}$, immersed in a salt solution of multipolar particles, at long distances. I assume that all the peripheral charges $q_{\alpha}$ have the same sign and the distribution functions are spherically symmetric ones, i.e. $\omega_{\alpha}(\bold{r})=\omega_{\alpha}(|\bold{r}|)$. In that case the characteristic functions can be expanded into a power series near $\bold{k}=0$, so that $\omega_{\alpha}(\bold{k})=1-k^2\left<\xi_{\alpha}^2\right>/2!+k^4\left<\xi_{\alpha}^4\right>/4!+O(k^6)$. Placing the test ion at the origin and taking into account that $\rho_{ext}(\bold{r})=q_0\delta(\bold{r})$, I obtain 
\begin{equation}
\label{psi_2}
\psi(\bold{r})\simeq \frac{4\pi q_0}{\varepsilon_{b}}\int\frac{d\bold{k}}{(2\pi)^3}\frac{e^{i\bold{k}\bold{r}}}{k^2+\tilde{\kappa}_s^2+L_{Q}^2k^4},
\end{equation}
where $\varepsilon_{b}=\varepsilon+4\pi n\sum_{\alpha}q_{\alpha}^2\left<\xi_{\alpha}^2\right>/(k_{B}T)$ is the bulk dielectric permittivity of the solution, taking into account the dielectric effect of the multipolar molecules, $\tilde{\kappa}_{s}=(8\pi e^2 I/k_{B}T\varepsilon_b)^{1/2}$ is the inverse Debye length of the salt ions in the dielectric medium with the permittivity $\varepsilon_b$. The $"$quadrupolar$"$ length $L_Q$ \cite{budkov2019Astatistical,slavchov2014quadrupole} is determined by the relation
\begin{equation}
L_{Q}^2=\frac{\pi n}{\varepsilon_b k_{B}T} \sum\limits_{\alpha,\gamma} q_{\alpha}q_{\gamma}\left(\left<\xi_{\alpha}^2\right>\left<\xi_{\gamma}^2\right>\left(1-\delta_{\alpha\gamma}\right)-\frac{1}{3}\delta_{\alpha\gamma}\left<\xi_{\alpha}^4\right>\right),
\end{equation}
where $\delta_{\alpha\gamma}$ is the Kronecker symbol. The quadrupolar length is positive at sufficiently large number $m$ of the identical peripheral charges $q_{\alpha}=-q/m$ with $\left<\xi_{\alpha}^2\right>=l^2/6$. In this case of pure quadrupolar molecules, the quadrupolar length is determined by the following expression \cite{budkov2019Astatistical}
\begin{equation}
L_{Q}^2\simeq\frac{\pi q^2nl^4}{9\varepsilon k_{B}T}.
\end{equation}
Therefore, for the case of quadrupolar molecules calculation of integral in eq. (\ref{psi_2}) yields \cite{budkov2019Astatistical}
\begin{equation}
\psi(\bold{r})=\frac{q_0}{\varepsilon_b r}\times
\begin{cases}
\frac{\exp\left(-\kappa_{-} r\right)-\exp\left(-\kappa_{+} r\right)}{\sqrt{1-4\tilde{\kappa}_{s}^2 L_{Q}^2}},& \tilde{\kappa}_{s} L_{Q}< \frac{1}{2}\,\\
\frac{r}{\sqrt{2}L_{Q}}\exp\left(-\frac{r}{\sqrt{2}L_Q}\right), &\tilde{\kappa}_{s} L_{Q}= \frac{1}{2}\,\\
\frac{\exp\left(-\kappa_0 r\right)}{\sqrt{4\tilde{\kappa}_{s}^2 L_{Q}^2-1}}\sin\left(\frac{r}{\lambda}\right),&\tilde{\kappa}_{s}L_{Q}> \frac{1}{2},
\end{cases}
\end{equation}
where $r=|\bold{r}|$ and the following short-hand notations
\begin{equation}
\kappa_{\pm}=\frac{\left(1\pm\sqrt{1-4\tilde{\kappa}_{s}^2L_{Q}^2}\right)^{1/2}}{\sqrt{2}L_{Q}},~~\kappa_0=\tilde{\kappa}_{s}\sqrt{1+1/(2\tilde{\kappa}_{s} L_{Q})},
\end{equation}
\begin{equation}
\lambda=\frac{1}{\tilde{\kappa}_s\sqrt{1-1/(2\tilde{\kappa}_{s} L_{Q})}}
\end{equation}
are introduced. As is seen, the electrostatic potential behaves at long distances in two qualitatively different manners. If the quadrupolar length is less than half of the Debye length, the potential monotonically decreases at long distances. However, in the case when the quadrupolar length exceeds half of the Debye length, the potential damps with oscillations, characterized by a certain wavelength $\lambda$ and a decrement $\kappa_0$. Such an asymptotic behavior of the potential was originally discovered by Slavchov \cite{slavchov2014quadrupole} in the framework of the pure phenomenological theory. I would like to note that the oscillations of the electrostatic potential is a peculiarity of the ion-quadrupole media. It is also interesting to note that analogous oscillations of the electrostatic potential can be observed in ionic liquids at electrified interfaces (so-called overscreening) \cite{bazant2011double,fedorov2014ionic}. Recently, using the idea that ions in ionic liquids can form electrically neutral quadrupolar clusters \cite{feng2019free,coles2019correlation}, Avni et al. \cite{avni2020charge} proposed a possible mechanism of the overscreening effect. I would like to stress that oscillations of the mean-force electrostatic potential in ionic media containing quadrupolar clusters is a new fundamental effect taking place in different ion-molecular systems. In terms of statistical mechanics, the obtained crossover from a monotonic decrease in the mean-force electrostatic potential to its oscillating decrease is related to exceeding of the so-called Fisher-Widom line \cite{fisher1969decay,stopper2019decay} on the phase diagram of the solution. Thus, the Fisher-Widom line for the ionic component of the solution is determined by the equation $\tilde{\kappa}_s L_{Q}=1/2$. The length $\kappa_0^{-1}$ characterizes the decay of the correlations of ionic concentration fluctuations, and is analogous to Ornstein-Zernike correlation length \cite{barrat2003basic}, while $\lambda$ is a measure of the $"$periodicity$"$ of positive ion-rich and negative ion-rich domains. \footnote{This is related to the fact that electrostatic potential $\psi(\bold{r})$ determines an asymptotic behavior of the pair correlation functions of ions.} The analogous charge periodicity is realized in salt solutions of weak polyelectrolytes \cite{borue1988statistical}. Note that similar periodicity of oil-rich and water-rich domains can be observed in microemulsions \cite{barrat2003basic}. I would like to note also that given analysis of electrostatic potential is valid only in case when $\tilde{\kappa}_s L_{Q}\sim 1$. In case when $\tilde{\kappa}_s L_{Q}\gg 1$ the low-k expansion of the screening function is not valid, and, thereby, higher order terms must be included.

\textbf{Electrostatic free energy.} 
The electrostatic free energy of the bulk solution is determined by eq. (\ref{el_free_en_lin_2}) with screening function (\ref{screen_func_3}). 
Now let me consider the case of identical peripheral charges $q_{\alpha}=-q/m$ with the same spherically symmetric distribution functions $\omega_{\alpha}(\bold{r})=\omega(|\bold{r}|)$. In this case, the screening function takes the form
\begin{equation}
\label{varkappa}
\varkappa^2(\bold{k})=\kappa_{s}^2+\frac{4\pi q^2n (m+1)}{\varepsilon k_{B}Tm}(1-\omega(\bold{k}))\left(1-\frac{m-1}{m+1}\omega(\bold{k})\right).
\end{equation}
For the case of dipolar molecules ($m=1$), eq. (\ref{varkappa}) yields the earlier obtained eq. (\ref{screen_func_2}). It is interesting to consider another limiting case $m\gg 1$, when the peripheral charges form a continuous charged cloud surrounding the central charge. In this case, the screening function takes the following form
\begin{equation}
\varkappa^2(\bold{k})\simeq\kappa_{s}^2+\frac{4\pi q^2n }{\varepsilon k_{B}T}(1-\omega(\bold{k}))^2.
\end{equation}

For the model distribution function determined by characteristic function (\ref{char_func_model}), one can obtain an analytical expression for the electrostatic free energy for the salt-free solution ($\kappa_s =0$), i.e.
\begin{equation}
\label{Fex}
F_{el}=-\frac{Vk_{B}T}{l^3}s(y),
\end{equation}
where $s(y)=\sqrt{6}\pi^{-1}\left((1+y)\sqrt{1+y/4}-{9y}/{8}-1\right)$, $y=2\pi q^2l^2n/(3\varepsilon k_{B}T)=\kappa^2 l^2/6$, and $\kappa=(4\pi q^2n/\varepsilon k_{B}T)^{1/2}$ is the inverse Debye length, attributed to the central charges. It is interesting to analyse the limiting cases, following from eq. (\ref{Fex}). Thus, I have
\begin{equation}
\frac{F_{el}}{Vk_BT}=
\begin{cases}
-\frac{5\sqrt{6}\pi q^4 l}{96 \left(\varepsilon k_{B}T\right)^2}n^2, &y\ll 1\,\\
-\frac{\kappa^3}{12\pi},&y\gg 1.
\end{cases}
\end{equation}
The first regime determines the case, when the effective size of charged cloud $l$ is much less than the Debye length $\kappa^{-1}$. In this case, the effective interactions of the multipolar molecules manifest themselves as the effective Van der Waals interaction  -- Kirkwood-Shumaker attractive interaction \cite{kirkwood1952influence,advzic2015charge,avni2019charge}, which is caused by the spatial fluctuations of the charged clouds of the molecules. In the opposite case, the electrostatic free energy is described by the DH limiting law. In this case, a salt-free solution can be described by the one-component plasma model (see papers \cite{brilliantov1998accurate,ortner1999equation} and report \cite{levin2002electrostatic}). Indeed, in the latter case the central charges are immersed in the background of collectivized charged clouds. In the regime $y\ll 1$, one can obtain \cite{budkov2019Astatistical}
\begin{equation}
\label{small_yc}
\frac{F_{el}}{V k_{B}T}=-\frac{\kappa_s^3}{12\pi}-\frac{\sqrt{6}\pi q^2 n}{\varepsilon k_{B}T l}h_{1}(y_s)-\frac{\sqrt{6}\pi q^4 l n^2}{96(\varepsilon k_{B}T)^2}h_2(y_s)+O(n^3),
\end{equation}
where $y_s=\kappa_s^2 l^2/6$ and the following functions are introduced
\begin{equation}
h_1(y_s)=\frac{y_s(1+2\sqrt{y_s})}{(1+\sqrt{y_s})^2},~h_2(y_s)=\frac{40y_s^{3/2}+8y_s^2+48y_s+25\sqrt{y_s}+5}{(1+\sqrt{y_s})^5}.
\end{equation}
In the regime when $y_{s}\gg 1$ and $y\gg 1$, I obtain
\begin{equation}
\label{DH_OCP}
\frac{F_{el}}{V k_{B}T}=-\frac{(\kappa^2+\kappa_s^2)^{3/2}}{12\pi}.
\end{equation}
The latter expression determines the DH limiting law for the one-component plasma in the presence of salt ions.

\textbf{Multipolar molecules with a hard core.}
Now I would like to demonstrate how one can take into account the excluded volume interactions between the multipolar molecules. For this purpose, let me assume that the central charges are placed at the centers of hard spheres with a diameter $d$. Let me also consider the peripheral charges as phantom ones, which can penetrate freely inside the hard cores of multipolar molecules. For simplicity, I consider only the case of salt-free solution of the multipolar molecules. \footnote{For the salt solutions of multipolar molecules it is necessary to take into account the excluded volume correlations between ions and multipolar molecules. In principle, it can be done within the formalism formulated in \cite{budkov2019statistical}}. The screening function for this case takes the following form (for details, see Appendix II)
\begin{equation}
\label{screen}
\varkappa^2(\bold{k})=\frac{4\pi n}{\varepsilon k_{B}T}\left(|q(\bold{k})|^2S(\bold{k})+\sum\limits_{\alpha=1}^{m}q_{\alpha}^2(1-|\omega_{\alpha}(\bold{k})|^2)\right),
\end{equation}
where the function $q(\bold{k})$ is determined by eq. (\ref{q}) and $S(\bold{k})$ is the structure factor of the hard spheres system determined above. The electrostatic free energy is determined by eq. (\ref{el_free_en_lin_2}) with screening function (\ref{screen}). Considering only the case of identical peripheral charges $q_{\alpha}=-q/m$ and the case of $m\gg 1$, one can get
\begin{equation}
\varkappa^2(\bold{k})\simeq \frac{4\pi q^2 n}{\varepsilon k_{B}T}(1-\omega(\bold{k}))^2S(\bold{k}).
\end{equation}
Further, using the discussed in section I $"$mean-field$"$ approximation $S(\bold{k})\approx S(0)=nk_{B}T\chi_{T}=z_{0}$, where $\chi_{T}$ is the isothermal compressibility of the hard spheres system, I arrive at the following expression for the electrostatic free energy
\begin{equation}
\label{Fex2}
F_{el}=-\frac{Vk_{B}T}{l^3}s(\tilde{y}),
\end{equation}
where $\tilde{y}=yz_{0}$ and the auxiliary function $s(y)$ has been determined above. 
The limiting regimes resulting from eq. (\ref{Fex2}) are
\begin{equation}
\frac{F_{el}}{Vk_BT}=
\begin{cases}
-\frac{5\sqrt{6}\pi q^4 l}{96 \left(\varepsilon k_{B}T\right)^2}z_0^2 n^2, &y\ll 1\,\\
-\frac{\kappa^3}{12\pi}z_{0}^{3/2},&y\gg 1.
\end{cases}
\end{equation}
In the Percus-Yevick approximation \cite{hansen1990theory} the compressibility factor is determined by the expression \cite{hansen1990theory} $z_0=(1-\eta)^4/(1+2\eta)^2 <1$.
Therefore, as in the case of zwitterions with a hard core, accounting for the excluded volume correlations of the multipolar molecules reduces the absolute value of electrostatic contribution to the total free energy of solution.

\section{Conclusions and future prospects}
In this paper, in the context of the modern statistical physics of ion-molecular systems, I have outlined my recent developments in the statistical field theory of salt solutions of zwitterions and multipolar molecules. I have formulated a general field theoretic formalism, going beyond the standard simple fluids field theory paradigm. The presented findings allow taking into account the complex internal electric structure of zwitterionic and multipolar molecules in the salt solution media. I have demonstrated how the formulated formalism can be applied to the calculations of various thermodynamic properties of the salt solutions of zwitterions and multipolar molecules. I believe that this formalism can be used in future as a theoretical background, making it possible to properly describe the electrostatic interactions within the theoretical models of ion-molecular systems, such as salt solutions of metal-organic complexes, complex colloids, phospholipids, betaines, proteins, {\sl etc.}

\section*{Acknowledgments}
The author has benefited from conversations with I.Ya. Erukhimovich, A.A. Kornyshev, N.V. Brilliantov, A.I. Victorov, M.G. Kiselev, A.L. Kolesnikov, and G.N. Chuev. The author thanks N.N. Kalikin for help with text preparation. The author thanks reviewers for valuable comments allowing to improve the text. The author is very grateful to Editorial Board for opportunity to present his theoretical results in the themed collection PCCP Emerging Investigator. The work was prepared within the framework of the Academic Fund Program at the National Research University Higher School of Economics in 2019-2021 (grant No 19-01-088) and by the Russian Academic Excellence Project $"$5-100$"$.    

\section{Appendix I: Electrostatic free energy of a solution of multipolar molecules}
In this Appendix, I would like to describe the process of deriving the expression for the electrostatic free energy of the bulk salt solution of the multipolar molecules within the RPA. As it was already mentioned in the main text, the configuration integral of a solution of multipolar molecules within the RPA takes the following general form
\begin{equation}
\label{RPA_gen}
Q\approx \exp\left\{-S[i\psi]+\frac{1}{2}tr\left(\ln \mathcal{G} -\ln G_{0}\right)\right\},
\end{equation}
where the symbol $tr(..)$ denotes a trace of the integral operator $A$ with the kernel $A(\bold{r},\bold{r}^{\prime})$ in accordance with the following definition
\begin{equation}
\label{tr}
tr(A)=\int d\bold{r}A(\bold{r},\bold{r}).
\end{equation}
The inverse Green function can be written as follows
\begin{equation}
\label{mult_green}
\mathcal{G}^{-1}(\bold r,\bold r'|\psi)=k_{B}T\frac{\delta^2 S[i\psi]}{\delta \varphi(\bold{r})\delta \varphi(\bold{r}')}=G_0^{-1}(\bold r,\bold r')+\mathcal{R}_{i}(\bold r,\bold r'|\psi)+\mathcal{R}_{h}(\bold r,\bold r'|\psi)
\end{equation}
where
\begin{equation}
\mathcal{R}_{i}(\bold{r},\bold{r}^{\prime}|\psi)=\beta e^2\sum\limits_{j=1}^{s}z_{j}^2c_{j}e^{-\beta z_{j}e\psi(\bold{r})}\delta(\bold r-\bold r'),
\end{equation}
and
\begin{equation}
\nonumber
\mathcal{R}_{h}(\bold{r},\bold{r}^{\prime}|\psi)=\beta n\sum\limits_{\delta,\gamma}q_{\delta}q_{\gamma}
\left<e^{-\beta\sum\limits_{\alpha=1}^{m}q_{\alpha}\left(\psi(\bold{r}+\bold{\xi}_{\alpha}-\bold\xi_{\gamma})-\psi(\bold{r}-\bold\xi_{\gamma})\right)}
\left(\delta(\bold{r}-\bold{r}^{\prime}+\bold{\xi}_{\delta}-\bold\xi_{\gamma})-\delta(\bold{r}-\bold{r}^{\prime}-\bold\xi_{\gamma})\right)\right>_{\xi}
\end{equation}
\begin{equation}
-\beta n\sum\limits_{\delta,\gamma}q_{\delta}q_{\gamma}
\left<e^{-\beta\sum\limits_{\alpha=1}^{m}q_{\alpha}\left(\psi(\bold{r}+\bold{\xi}_{\alpha})-\psi(\bold{r})\right)}
\left(\delta\left(\bold{r}-\bold{r}^{\prime}+\bold{\xi}_{\delta}\right)-\delta\left(\bold{r}-\bold{r}^{\prime}\right)\right)\right>_{\xi}.
\end{equation}
I would like to note that for the case of dipolar molecules ($m=1$) eq. (\ref{mult_green}) transforms into eq. (\ref{dip_green}) written above. In the absence of the external charges (i.e., $\rho_{ext}(\bold{r})=0$), the self-consistent field potential $\psi(\bold{r})=0$, so that the mean-field contribution $F_{el}^{(MF)}=k_{B}T S[0]=0$. Therefore, the inverse Green function takes the following simplified form:
\begin{equation}
\nonumber
\mathcal{G}^{-1}(\bold r,\bold r'|0)=G^{-1}(\bold r,\bold r')=G_0^{-1}(\bold r,\bold r')+R_{i}(\bold{r},\bold{r}^{\prime})+R_{h}(\bold{r},\bold{r}^{\prime})
\end{equation}
where the following kernels
\begin{equation}
R_{i}(\bold{r},\bold{r}^{\prime})=\mathcal{R}_{i}(\bold{r},\bold{r}^{\prime}|0)=2 \beta I e^2 \delta(\bold r-\bold r'),
\end{equation}
\begin{equation}
R_{h}(\bold{r},\bold{r}^{\prime})=\mathcal{R}_{h}(\bold{r},\bold{r}^{\prime}|0)
=\beta n\sum\limits_{\delta,\gamma}q_{\delta}q_{\gamma}\left<\delta\left(\bold{r}-\bold r'+\bold{\xi}_{\delta}-\bold\xi_{\gamma}\right)-2\delta\left(\bold{r}-\bold r'-\bold\xi_{\gamma}\right)+\delta\left(\bold{r}-\bold r'\right)\right>_{\xi}
\end{equation}
are introduced; $I$ is the ionic strength of the solution determined in the main text.

The electrostatic free energy, thus, is determined by
\begin{equation}
\label{F_el_simp}
F_{el}\approx \frac{k_{B}T}{2}tr\left(\ln G_0 - \ln G\right)=\frac{Vk_{B}T}{2}\int\frac{d\bold k}{(2\pi)^3}\ln\frac{G_0(\bold k)}{G(\bold k)},
\end{equation}
where $G_0(\bold k)=4\pi/(\varepsilon k^2)$ and $G(\bold k)=4\pi/(\varepsilon(k^2+\varkappa^2(\bold k)))$ are the Fourier-images of the Green functions; the screening function $\varkappa^2(\bold{k})$ is determined by eq. (\ref{screen_func_3}).

Now let me derive eq. (\ref{F_el_simp}). Thus, I have
\begin{equation}
tr\left(\ln G_0 - \ln G\right)=tr\ln(G^{-1}G_0)=tr\ln((G_{0}^{-1}+R)G_{0})=tr\ln(\bold{I}+A),
\end{equation}
where $\bold{I}$ is the identity operator and $A=RG_0$ with $R=R_{i}+R_h$. Therefore, I obtain
\begin{equation}
\label{trace}
tr\ln(\bold{I}+A)=\sum\limits_{n=1}^{\infty}\frac{(-1)^{n+1}}{n}tr(A^{n}).
\end{equation}
In accordance with the above-mentioned definition (\ref{tr}), I can write
\begin{equation}
tr(A^{n})=\int d\bold{r} (A^{n})(\bold{r},\bold{r}),
\end{equation}
where
\begin{equation}
(A^{n})(\bold{r},\bold{r}^{\prime})=\int d\bold{r}_1..\int d\bold{r}_{n-1}A(\bold{r}-\bold{r}_1)A(\bold{r}_1-\bold{r}_2)...A(\bold{r}_{n-2}-\bold{r}_{n-1})A(\bold{r}_{n-1}-\bold{r}^{\prime})\nonumber
\end{equation}
\begin{equation}
=\int\frac{d\bold{k}}{(2\pi)^3}e^{i\bold{k}(\bold{r}-\bold{r}^{\prime})}\tilde{A}^{n}(\bold{k}),
\end{equation}
where 
\begin{equation}
\tilde{A}(\bold{k})=\int d\bold{r}e^{-i\bold{k}\bold{r}}A(\bold{r})
\end{equation}
is the Fourier-image of the function $A(\bold{r})$. Therefore, I obtain
\begin{equation}
tr(A^n)=V\int\frac{d\bold{k}}{(2\pi)^3}\tilde{A}^{n}(\bold{k}),
\end{equation}
where I take into account that $\int d\bold{r} =V$.
Further, using eq. (\ref{trace}), I arrive at the following equality
\begin{equation}
tr\ln(\bold{I}+A)=V\int\frac{d\bold{k}}{(2\pi)^3}\ln\left(1+\tilde{A}(\bold{k})\right).
\end{equation}
Therefore, taking into account that $\tilde{A}(\bold{k})=\tilde{R}(\bold{k})G_{0}(\bold{k})=\varkappa^{2}(\bold{k})/k^2$, I finally obtain
\begin{equation}
F_{el}=\frac{Vk_{B}T}{2}\int\frac{d\bold k}{(2\pi)^3}\ln\left(1+\frac{\varkappa^{2}(\bold{k})}{k^2}\right)=\frac{Vk_{B}T}{2}\int\frac{d\bold k}{(2\pi)^3}\ln\frac{G_0(\bold k)}{G(\bold k)},
\end{equation}
where I take into account that
\begin{equation}
\tilde{R}(\bold{k})=\tilde{R}_{i}(\bold{k})+\tilde{R}_{h}(\bold{k})=2\beta I+\beta n\left(|q(\bold{k})|^2+\sum\limits_{\alpha=1}^{m}q_{\alpha}^2(1-|\omega_{\alpha}(\bold{k})|^2)\right).
\end{equation}

Subtracting the electrostatic self-energy $E_{self}=V/(2 (2\pi)^{3})\int {d\bold{k}}\varkappa^2(\bold{k})/k^2$ of the particles, I arrive at eq. (\ref{el_free_en_lin_2}).

\section{Appendix II: Derivation of electrostatic free energy of a salt-free solution of multipolar molecules with excluded volume interactions}
In this appendix, I consider the derivation of the electrostatic free energy of a salt-free solution of multipolar molecules taking into account the excluded volume interactions between them. As it was already mentioned in the main text, the central charge $q$ is placed at the center of a hard sphere with a diameter $d$, whereas the peripheral charges $q_{\alpha}$ ($\alpha=1,...,m$) are the phantom ones which can freely penetrate inside the hard core. Therefore, the configuration integral of such a system can be written as follows
\begin{equation}
Q=\int d\Gamma_{h}\exp\left[-\beta H_{hc}-\beta H_{el}\right],
\end{equation}
where the integration over the configurations of the multipolar molecules is performed in accordance with definitions (\ref{dGamma_h}) and (\ref{dGamma_j}). The first term in the integrand exponent describes the contribution of the excluded volume interactions between the multipolar molecules, i.e.
\begin{equation}
H_{hc}=\frac{1}{2}\sum\limits_{i\neq j}U_{hc}(\bold{R}_{i}-\bold{R}_{j}),
\end{equation}
where 
\begin{equation}
U_{hc}(\bold{r})=
\begin{cases}
0,|\bold{r}|>0\,\\
\infty,|\bold{r}|\leq 0,
\end{cases}
\end{equation}
is the hard core potential. The second term describes the potential energy of electrostatic interactions of the multipolar molecules, i.e.
\begin{equation}
H_{el}=\frac{1}{2}\left(\hat{\rho}_{h}G_{0}\hat{\rho}_{h}\right)=\frac{1}{2}\int d\bold{r}\int d\bold{r}^{\prime}\hat{\rho}_{h}(\bold{r})G_{0}(\bold{r}-\bold{r}^{\prime})\hat{\rho}_{h}(\bold{r}^{\prime}),    
\end{equation}
where $\hat{\rho}_{h}(\bold{r})$ is the microscopic charge density of multipolar molecules which is determined by eq. (\ref{rho_h}). Further, using the hard spheres system as a reference system, I can recast the configuration integral in the thermodynamic perturbation theory fashion, i.e. 
\begin{equation}
Q=Q_{hs}\left<\exp\left[-\frac{1}{2}\left(\hat{\rho}_{h}G_{0}\hat{\rho}_{h}\right)\right]\right>,
\end{equation}
where 
\begin{equation}
Q_{hs}=\int d\Gamma_{h}\exp\left[-\beta H_{hc}\right]   
\end{equation}
is the configuration integral of the reference system and the notation
\begin{equation}
\left<(...)\right>=\frac{1}{Q_{hs}}\int d\Gamma_{h}\exp\left[-\beta H_{hc}\right](...)
\end{equation}
denotes the averaging over the microstates of the reference system. Using the Hubbard-Stratonovich transformation, I arrive at the following representation of the configuration integral
\begin{equation}
Q=Q_{hs}\int\frac{\mathcal{D}\varphi}{C}\exp\left[-\frac{\beta}{2}(\varphi G_{0}^{-1}\varphi)\right]\left<\exp\left[i\beta (\hat{\rho}_{h}\varphi)\right]\right>.
\end{equation}
The cumulant expansion \cite{kubo1962generalized} gives
\begin{equation}
\left<\exp\left[i\beta (\hat{\rho}_{h}\varphi)\right]\right>=
\exp\left[i\beta (\left<\hat{\rho}_{h}\right>\varphi)-\frac{\beta^2}{2}\left(\varphi \mathcal{K}\varphi\right)+...\right],
\end{equation}
where 
\begin{equation}
\mathcal{K}(\bold{r}-\bold{r}^{\prime})=\left<\hat{\rho}_{h}(\bold{r})\hat{\rho}_{h}(\bold{r}^{\prime})\right>-\left<\hat{\rho}_{h}(\bold{r})\right>\left<\hat{\rho}_{h}(\bold{r}^{\prime})\right>
\end{equation}
is the charge density correlation function obtaned by averaging over the microstates of the hard spheres system. Taking into account that $\left<\hat{\rho}_{h}(\bold{r})\right>=0$, we arrive at 
\begin{equation}
\nonumber
\mathcal{K}(\bold{r}-\bold{r}^{\prime})=\left<\hat{\rho}_{h}(\bold{r})\hat{\rho}_{h}(\bold{r}^{\prime})\right> 
\end{equation}
\begin{equation}
\nonumber
=\sum\limits_{i,j}\sum\limits_{\alpha,\gamma}q_{\alpha}q_{\gamma}\left<\left(\delta\left(\bold{r}-\bold{R}_{i}-\bold{\xi}_{i}^{(\alpha)}\right)-\delta\left(\bold{r}-\bold{R}_{i}\right)\right)\left(\delta\left(\bold{r}-\bold{R}_{j}-\bold{\xi}_{j}^{(\gamma)}\right)-\delta\left(\bold{r}-\bold{R}_{j}\right)\right)\right>
\end{equation}
\begin{equation}
=\sum\limits_{i,j}\sum\limits_{\alpha,\gamma}q_{\alpha}q_{\gamma}\int\frac{d\bold{k}}{(2\pi)^3}\int\frac{d\bold{q}}{(2\pi)^3}e^{i\bold{k}\bold{r}+i\bold{q}\bold{r}^{\prime}}\left<e^{-i\bold{k}\bold{R}_{i}-i\bold{q}\bold{R}_{j}}\right>_{hs}\left<\left(e^{-i\bold{k}\bold{\xi}_{i}^{(\alpha)}}-1\right)\left(e^{-i\bold{q}\bold{\xi}_{j}^{(\gamma)}}-1\right)\right>_{\xi},
\end{equation}
where I used the Fourier-representation of the delta-function
\begin{equation}
\delta(\bold{x})=\int\frac{d\bold{k}}{(2\pi)^3}e^{i\bold{k}\bold{x}}  
\end{equation}
and extracted the averaging $\left<(...)\right>_{hs}$ over the microstates of the hard-spheres and the average $\left<(...)\right>_{\xi}$ over the random variables $\bold{\xi}_{i}^{(\alpha)}$. Further, taking into account that in the thermodynamic limit \cite{Gray}
\begin{equation}
\left<e^{-i\bold{k}\bold{R}_{i}-i\bold{q}\bold{R}_{j}}\right>_{hs}=\frac{\left(2\pi\right)^3}{V}\delta(\bold{k}+\bold{q})\left<e^{-i\bold{k}(\bold{R}_{i}-\bold{R}_{j})}\right>_{hs}
\end{equation}
and using the definition of the structure factor \cite{hansen1990theory,Gray}
\begin{equation}
S(\bold{k})=\frac{1}{N}\sum\limits_{i,j}\left<e^{-i\bold{k}(\bold{R}_{i}-\bold{R}_{j})}\right>_{hs}
\end{equation}
after some algebra, I obtain
\begin{equation}
\mathcal{K}(\bold{r}-\bold{r}^{\prime})=n\int\frac{d\bold{k}}{(2\pi)^3}e^{i\bold{k}(\bold{r}-\bold{r}^{\prime})}\left(|q(\bold{k})|^2S(\bold{k})+\sum\limits_{\alpha=1}^{m}q_{\alpha}^2(1-|\omega_{\alpha}(\bold{k})|^2)\right),
\end{equation}
where I take into account that $\omega_{\alpha}(\bold{k})=\left<e^{-i\bold{k}\bold{\xi}_{i}^{(\alpha)}}\right>_{\xi}$ and the auxiliary function $q(\bold{k})$ is determined by eq. (\ref{q}).

Therefore, I arrive at the following expression for the configuration integral in the RPA 
\begin{equation}
Q=Q_{hs}\int\frac{\mathcal{D}\varphi}{C} \exp\left[-\frac{\beta}{2}(\varphi G^{-1}\varphi)\right],
\end{equation}
where the renormalized inverse Green function 
\begin{equation}
\label{Green_hs}
G^{-1}(\bold{r},\bold{r}^{\prime})=G_{0}^{-1}(\bold{r},\bold{r}^{\prime})+\beta \mathcal{K}(\bold{r}-\bold{r}^{\prime})
\end{equation}
is introduced.
Eq. (\ref{Green_hs}) can be rewritten in the Fourier-representation as follows
\begin{equation}
G(\bold{k})=\frac{G_{0}(\bold{k})}{1+\beta \tilde{\mathcal{K}}(\bold{k})G_{0}(\bold{k})}=\frac{4\pi}{\varepsilon(k^2+\varkappa^2(\bold{k}))},
\end{equation}
where the screening function is determined by the following expression
\begin{equation}
\label{scr}
\varkappa^2(\bold{k})=\frac{4\pi n}{\varepsilon k_{B}T}\left(|q(\bold{k})|^2S(\bold{k})+\sum\limits_{\alpha=1}^{m}q_{\alpha}^2(1-|\omega_{\alpha}(\bold{k})|^2)\right).
\end{equation}
Note that for the case of zwitterionic molecules ($m=1$) with a hard core, screening function (\ref{scr}) transforms into eq. (\ref{varkappa2}). I would also like to note that for the structure factor $S(\bold{k})$ of the hard spheres system I can use the analytical expression based on the Percus-Yevick approximation (see, for instance, \cite{hansen1990theory,santos2016exact}).

The excess free energy of the solution in the RPA takes the form
\begin{equation}
F_{ex}=-k_{B}T\ln{Q}=F_{hs}+\frac{Vk_{B}T}{2}\int\frac{d\bold{k}}{(2\pi)^3}\ln\frac{G_{0}(\bold{k})}{G(\bold{k})}.
\end{equation}
Subtracting the electrostatic self-energy from the final expression, I obtain
\begin{equation}
F_{ex}=F_{hs}+\frac{V k_{B}T}{2}\int\frac{d\bold{k}}{(2\pi)^3}\left(\ln\left(1+\frac{\varkappa^2(\bold{k})}{k^2}\right)-\frac{\varkappa^2(\bold{k})}{k^2}\right).
\end{equation}
For the excess free energy $F_{hs}$ of the hard spheres system I can use the Percus-Yevick or the Carnahan-Starling approximations \cite{hansen1990theory,santos2016exact}.

\bibliography{name}

\end{document}